\documentclass[reprint,aps,prl,amssymb]{revtex4-2}

\usepackage{graphicx}
\usepackage{amsmath}
\usepackage[svgnames]{xcolor} 
\usepackage[colorlinks=True, linkcolor=SlateBlue,
            citecolor=SteelBlue,urlcolor=BlueViolet]{hyperref}
\usepackage[capitalise]{cleveref}

\usepackage{aas_macros}

\newcommand{\imag}{\text{i}}
\newcommand{\out}{\text{o}}

\newcommand{\vect}[1]{\boldsymbol{#1}}
\newcommand{\uvect}[1]{\vect{\hat{#1}}}

\begin{document}

\author{Hang Yu}\thanks{hangyu@caltech.edu}
\affiliation{TAPIR, Walter Burke Institute for Theoretical Physics, Mailcode 350-17 California Institute of Technology, Pasadena, CA 91125, USA}

\author{Yanbei Chen}
\affiliation{TAPIR, Walter Burke Institute for Theoretical Physics, Mailcode 350-17 California Institute of Technology, Pasadena, CA 91125, USA}

\title{Direct determination of supermassive black hole properties with gravitational-wave radiation from surrounding stellar-mass black hole binaries}

\begin{abstract}
A significant number of stellar-mass black-hole (BH) binaries may merge in galactic nuclei or in the surrounding gas disks. With purposed space-borne gravitational-wave observatories, we may use such a binary as a signal carrier to probe modulations induced by a central supermassive BH (SMBH), which further allows us to place constraints on the SMBH's properties. We show in particular the de Sitter precession of the inner stellar-mass binary's orbital angular momentum (AM) around the AM of the outer orbit will be detectable if the precession period is comparable to the duration of observation, typically a few years. Once detected, the precession can be combined with the Doppler shift arising from the outer orbital motion to determine the mass of the SMBH and the outer orbital separation individually and each with percent-level accuracy. If we further assume a joint detection by space-borne and ground-based detectors, 
the detectability threshold could be extended to a precession period of $\sim 100\,{\rm yr}$. 
\end{abstract}

\maketitle


\emph{Introduction} -- A significant number of stellar-mass binary black holes (BH) detectable by LIGO~\cite{LSC:15} and Virgo~\cite{Acernese:15} may merge in the vicinity of supermassive BHs (SMBHs) due to both dynamical interactions~\cite{OLeary:09, Antonini:12, Antonini:16, Petrovich:17, Leigh:18,Chen:18,Fragione:19} and gaseous effects if accretion disks are present~\cite{McKernan:12, Bartos:17,Stone:17, McKernan:18, Tagawa:19, McKernan:19, Yang:19, Secunda:19}. This possibility is strengthened as the Zwicky Transient Facility~\cite{Bellm:19, Graham:19} detected a potential electromagnetic counterpart~\cite{Graham:20} to the LIGO-Virgo event GW190521~\cite{GW190521a, GW190521b},  consistent with a binary BH merger in the accretion disk of an active galactic nucleus (AGN). 

Beyond ground-based detectors, multiple space-borne gravitational-wave (GW) observatories have been planned/conceived for the coming decades, including LISA~\cite{Amaro-Seoane:17}, TianQin~\cite{Luo:16}, Taiji~\cite{Hu:17}, B-DECIGO~\cite{Nakamura:16, Kawamura:20}, Decihertz Observatories~\cite{Sedda:19},  and TianGO~\cite{Kuns:19}. Their sensitivities cover the 0.001-1 Hz band where a typical stellar-mass BH binary stays in band for years. It thus opens up the possibility of using a stellar-mass BH binary as a carrier to probe modulations induced by a tertiary perturber which, as argued above, can be an SMBH in many cases. This is in analog to how pulsars are used to test the strong-field relativity~\cite{Damour:92} and it offers a complementary way to probe SMBH properties to extreme and very extreme mass-ratio inspirals (EMRI and X-MRI)~\cite{Amaro-Seoane:18, Amaro-Seoane:19, Han:19, Han:20}. 

The leading-order modulation is a Doppler shift due to the inner binary's orbital motion around the SMBH\footnote{For future convenience, we will refer to the orbit of the stellar-mass BH binary as the inner orbit, and its center of mass orbiting the SMBH as the outer orbit.}, creating frequency sidebands at $\Omega_\out = (M_3/a_\out^3)^{1/2}$ with $M_3$ the mass of the SMBH and $a_\out$ the semi-major axis of the outer orbit. 
The extra dephasing of this effect can be determined up to $a_\out\simeq 1\,{\rm pc}$~\cite{Inayoshi:17}. When $2\pi/\Omega_\out \sim T_{\rm obs}$ with $T_{\rm obs}$ the duration of observation, $\Omega_\out$ can be further resolved to constrain the mass density enclosed by the outer orbit~\cite{Randall:19}. 

In the Letter, we extend the field by including higher-order effects.
The most significant one is that the inner orbital angular momentum (AM) $\vect{L}_\imag$ will experience a de Sitter-like (dS) precession around the outer AM $\vect{L}_\out$ whose secular effect is~\cite{Will:18, Liu:19, Yu:20b}
\begin{equation}
    \frac{d\uvect{L}_\imag}{dt} = \Omega_{\rm dS} \uvect{L}_\out \times \uvect{L}_\imag = \frac{3}{2}\frac{M_3}{a_\out (1-e_\out^2)}\Omega_\out\uvect{L}_\out \times \uvect{L}_\imag,\label{eq:dS_prec}
\end{equation}
where $e_\out$ is the eccentricity of the outer orbit. Here we have used the hat symbol to indicate unity vectors. As the binary precesses, the waveform undergoes both amplitude and phase modulations, thereby allowing the extraction of the preccession signatures. 

\begin{figure}
  \centering
  \includegraphics[width=0.85\columnwidth]{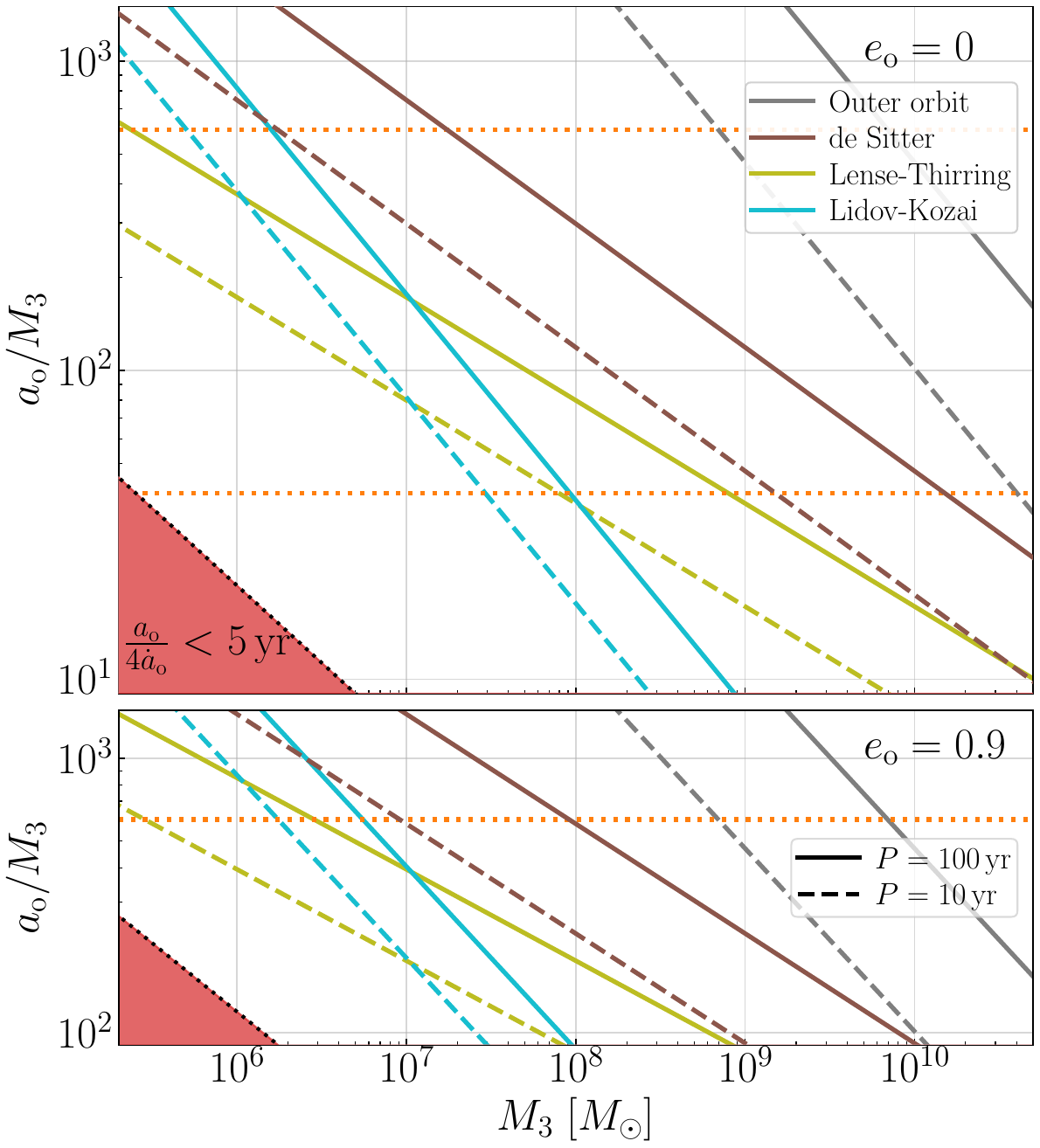}
  \caption{Periods of various dynamical processes. Approximately, below the solid trace an effect is detectable with joint detection by space-borne and ground-based detectors. Below the dashed trace, the effect can be constrained with a single TianGO-like detector.  We set the lower y-limit in each panel to $a_\out(1-e_\out)=9 M_3$. The decay of the outer orbit and hence the shaded region with $a_\out/4\dot{a}_\out<5\,{\rm yr}$ is discarded. The triple stability~\cite{Kiseleva:96} is always satisfied in the upper panel and its boundary is similar to the shaded region in the lower one. }
\label{fig:periods}
\end{figure}

We illustrate the periods of the dS precession in the $(M_3,\, a_\out)$ space in Fig.~\ref{fig:periods} with brown traces. The upper panel assumes a circular outer binary and the lower one has $e_\out=0.9$. The solid (dashed) traces correspond to $P_{\rm dS}=2\pi/\Omega_{\rm dS}=100\,(10)\,{\rm yr}$. As we will see later, these periods are the approximate detectability thresholds assuming a detection of a source 1\,Gpc away performed jointly by space-borne and ground-based detectors and by a TianGO-like detector alone. Also shown are the periods of the outer orbit (grey traces) and sub-leading corrections due to the Lense-Thirring precession (olive traces) and the Lidov-Kozai effect (i.e., the Newtonian tidal effect; cyan traces). The explicit expressions are provided in the Supplemental Material.

To connect to astrophysical formation mechanisms of the inner binary, we indicate in dotted-orange lines the expected locations of migration traps in accretion disks~\cite{Bellovary:16} where massive objects are likely to accumulate and binaries may frequently merge. 
We find $P_{\rm dS}<100\,{\rm yr}\ (10\,{\rm yr})$ at the migration trap at $\simeq 600\,M_3$ if $M_3\lesssim 2\times 10^7 \,M_\odot\ (2\times10^6\,M_\odot)$ and the outer orbit is circular. When the outer orbit is eccentric, the $P_{\rm dS}<100\,{\rm yr}$ boundary could be extended to further include $M_3\simeq 10^8\,{\rm yr}$. 

For bare nuclei, binaries can also be produced by various dynamical processes. Studies suggested a detection rate of $\mathcal{O}(10-100)\,{\rm yr^{-1}}$ BH binaries produced in the $a_\out \lesssim 0.1\,{\rm pc}$ region by the interaction channel~\cite{OLeary:09,VanLandingham:16,Antonini:16,Petrovich:17}. Assuming a density profile $\propto a_\out^{-2}$~\cite{OLeary:09}, it indicates $\mathcal{O}(0.1-1)$ detection per year in the central 0.001 pc region ($\simeq 200\,M_3$ for $M_3=10^8\,M_\odot$) where the dS precession could be significant. In fact, binary formed in this channel may be launched to an outer orbit with significant eccentricity that reduces $P_{\rm dS}$ by a factor $(1-e_\out^2)$ and allows binary formed at greater $a_\out$ to also experience significant precession (see the bottom panel of Fig.~\ref{fig:periods}).

Once observed, the dS precession allows a direct determination of properties of the SMBH and the outer orbit. 
Note its rate is $\Omega_{\rm dS}/\Omega_{\out}=M_3/\left[a_\out(1-e_\out^2)\right]$. When combined with the outer orbit's Doppler shift which tells us $\Omega_\out=\sqrt{M_3/a_\out^3}$ (and $e_\out$ for elliptical orbits as we illustrate in the Supplemental Material which includes Ref.~\cite{Flanagan:98}), we can therefore infer the values of $M_3$, $a_\out$, and $e_\out$ individually. 

Before this method, there are two common approaches to \emph{directly} determine the mass of an SMBH with a typical accuracy of tens of percent, either through directly observing the dynamics of star or gas around the SMBH, or through reverberation mapping of the continuum emission of AGNs~\cite{Peterson:14}. The former is limited to nearby ($\lesssim 100\,{\rm Mpc}$) SMBH and the later is applicable only to Type I (broad emission-line) AGNs, a trace of the population~\cite{Peterson:14}. 
LISA could also constrain SMBH masses via equal-mass inspirals and EMRIs. However, it is only sensitive to mergers with masses $\lesssim10^{7}\,M_\odot$~\cite{Hughes:02, Gair:04, Gair:17}. 
Our approach, on the other hand, probes SMBHs across almost the entire mass range to a distance of a few Gpc and applies independent of the SMBH being active or quiescent. It is thus an invaluable complimentary to the existing methods. Furthermore, it also determines the outer orbit 
via measuring $a_\out$ and $e_\out$ that are hard to be extracted otherwise at $\mathcal{O}({\rm Gpc})$ distances, thereby constraining the nuclei dynamics which currently has considerable theoretical uncertainties.

Hereafter, we will focus on the dS precession and how we can utilize it to measure $M_3$ and $a_\out$. We neglect the sub-leading Lense-Thirring precession and Lidov-Kozai oscillations for simplicity (but see Ref.~\cite{Fang:19}) and treat both the inner and outer orbits to be circular (we will discuss the effects of eccentricities at the end of the Letter). Gaseous frictions~\cite{Ostriker:99, Baruteau:11, Bartos:17, Antoni:19, Chen:20, Derdzinski:20} and encounters with background objects~\cite{Gultekin:04, Antonini:16} have characteristic timescales ranging from thousands to millions of years and therefore can be ignored over an observation over $T_{\rm obs}\simeq 5\,{\rm yr}$ (see the Supplemental Material for details).  All the parameters in the Letter correspond to their inferred values in the detector frame~\footnote{One can show that from the source frame to the detector frame, $M_3\to(1+z)M_3$ and $a_\out \to (1+z)a_\out$ due to a constant redshift $z$. Note $z$ may include both the cosmological redshift ($\sim 0.2$ at $1\,{\rm GPc}$) and that due to the gravitational potential of $M_3$ ($\lesssim 0.01$ for typical sources at $a_\out \gtrsim 100 M_3$).}. We use geometrical units $G=c=1$.


\begin{figure}
  \centering
  \includegraphics[width=0.92\columnwidth]{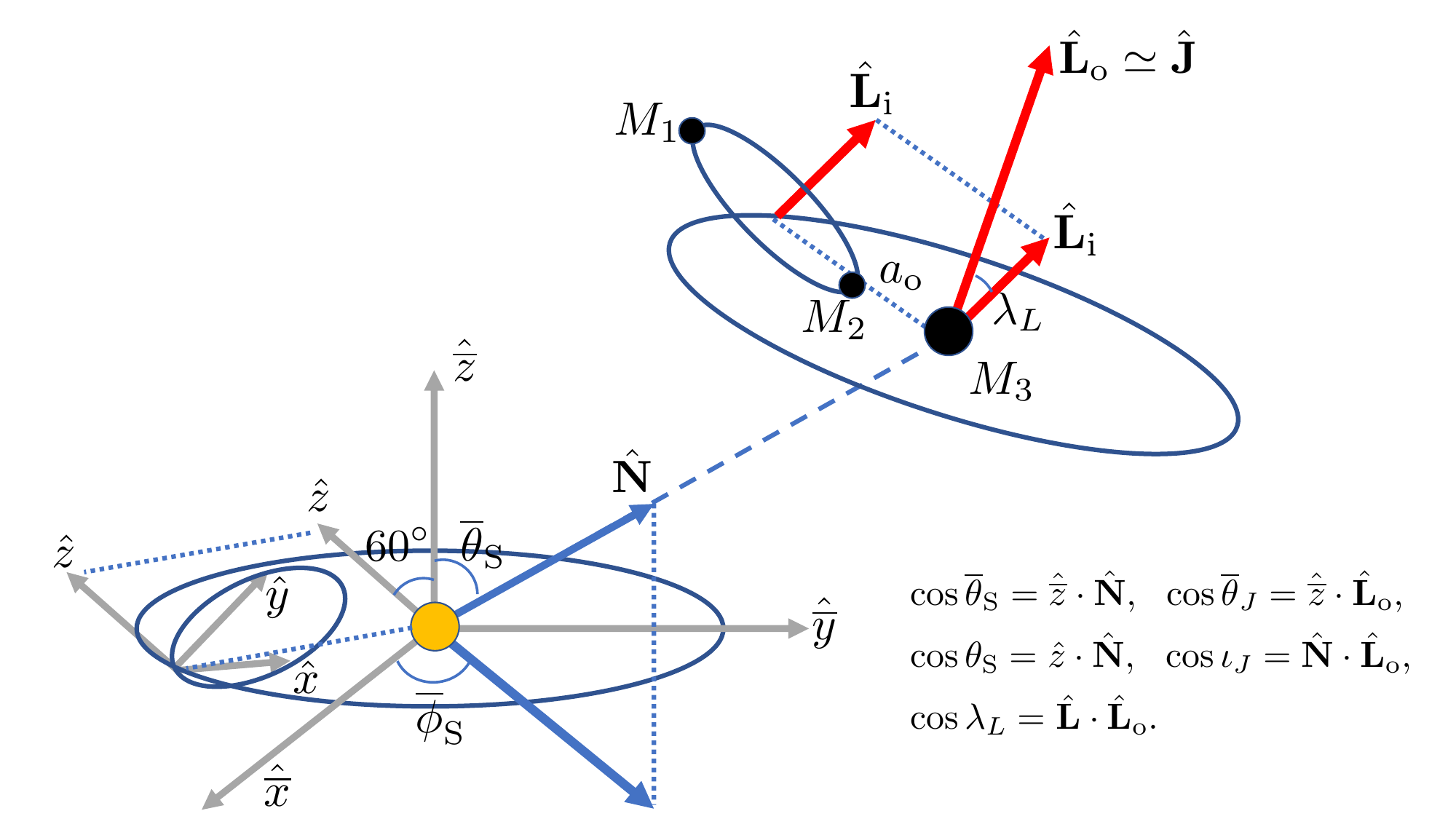}
  \caption{Cartoon illustrating the geometry of the problem. Note the amplitudes of vectors is chosen only for visualization purpose.}
\label{fig:cartoon}
\end{figure}

\emph{Waveforms} -- 
In Fig.~\ref{fig:cartoon} we demonstrate the geometry of the problem. We construct two reference frames. The $(x, y, z)$ frame is centered on the corner detector with $\uvect{x}$ and  $\uvect{y}$ pointing along two arms of TianGO~\cite{Kuns:19} (for LISA, this frame is constructed as in Ref.~\cite{Cutler:98}). As the detector frame changes in both location and orientation, we also construct a fixed solar frame $(\overline{x}, \overline{y}, \overline{z})$ with $\uvect{\overline{z}}$ perpendicular to the ecliptic. In the solar frame, the source's sky location $\uvect{N}$ and the total AM $\uvect{J}$ with $\vect{J}\equiv\vect{L}_\imag + \vect{L}_\out\simeq\vect{L}_\out$ are labeled with polar coordinates $(\overline{\theta}_{\rm S}, \overline{\phi}_{\rm S})$ and $(\overline{\theta}_{J}, \overline{\phi}_{J})$, respectively. We further define $\iota_J$ as the angle between $\uvect{N}$ and $\uvect{L}_\out$ and $\lambda_L$ the angle between $\uvect{L}_\imag$ and $\uvect{L}_\out$. The problem now becomes projecting the GW radiation characterized by a time-varying orientation $\uvect{L}_\imag(t)$ onto an antenna with also time-varying coordinates $(\uvect{x}, \uvect{y}, \uvect{z})$. 

To obtain the response, we follow Refs.~\cite{Apostolatos:94, Cutler:98}. The explicit expressions for various quantities could be found in the Supplemental Material.  The frequency-domain waveform under the stationary-phase approximation is 
\begin{align}
    \tilde{h}(f) &= \Lambda(f)\tilde{h}_{\rm C}(f)= [A_+^2(t)F_+^2(t) + A_\times^2(t)F_\times^2(t)]^{1/2} \nonumber \\ 
    &\times \exp\left\{-i\left[\Phi_{\rm p}(t)+ 2\Phi_{\rm T}(t) + \Phi_{\rm D}(t) \right] \right\} \tilde{h}_{\rm C}(f),
\end{align}
where $\Lambda$ characterizes the modulation due to antenna response and  $\tilde{h}_{\rm C}$ is the antenna-independent ``carrier''. We approximate $\tilde{h}_{\rm C}$ with the quadrupole formula, including four \emph{intrinsic} parameters, $(\mathcal{M}, D_{\rm L}, t_{\rm c}, \phi_{\rm c})$, corresponding to the chirp mass, luminosity distance, and time and phase of coalescence.  The antenna pattern depends on time which is further a function of frequency,
$ t(f) = t_{\rm c} - 5(8\pi f)^{-8/3}\mathcal{M}^{-5/3}. $

The changing orientations affects the amplitude both via $A_+=1+(\uvect{L}_\imag\cdot\uvect{N})^2$ and $A_\times = -2\uvect{L}_\imag\cdot\uvect{N}$, and via $F_{+(\times)}(\theta_{\rm S}, \phi_{\rm S}, \psi_{\rm S})$, where $\left(\theta_{\rm S}, \phi_{\rm S}\right)$ are the polar coordinates of $\uvect{N}$ in the $(x, y, z)$ frame and $\psi_{\rm S}$ is the polarization angle.

Besides amplitude modulations, there are also extra phase terms. 
The $\Phi_{\rm p}$ term characterizes the polarization phase, and 
the precession of $\uvect{L}_\imag$ further gives rise to a Thomas precession term $\Phi_{\rm T}$~\cite{Apostolatos:94}.
Lastly, $\Phi_{\rm D}$ describes a Doppler phase due to motions of both the outer orbit and the detector orbiting around the Sun. 

To this point the expressions are generic. A waveform is specified when one supplies information about the orbits (for $\Phi_{\rm D}$) and the orientations $\uvect{L}_\imag$ and $(\uvect{x}, \uvect{y}, \uvect{z})$. 

We model the Doppler phase as~\footnote{Here we have dropped the Shapiro time delay for simplicity, which contributes an extra phase $2\pi t_{\rm S}$ with $t_{\rm S}=2M_3\log[1/(1-\sin\iota_{J}\sin \phi_\out)]$~\cite{Blandford:76}. For most $\iota_J\neq 90^\circ$, $t_{\rm S}\sim 2M_3\ll a_\out$.}
\begin{align}
    \Phi_{\rm D} &= 2\pi f \left[a_\out \sin\iota_J\cos\left(\Omega_\out t - \phi^{(0)}\right) \right. \nonumber \\
    &+\left.{\rm AU}\sin\overline{\theta}_{\rm S} \cos\left(2\pi t/{\rm yr}- \overline{\phi}_{\rm S}\right)\right], 
    \label{eq:Phi_D}
\end{align}
where $\phi^{(0)}$ characterizes an initial phase for the outer orbit. The dS precession of $\uvect{L}_\imag$ around $\uvect{L}_\out$ can be written in terms of three additional parameters $(P_{\rm dS}, \lambda_{L}, \alpha_0)$ with $\alpha_0$ an initial phase characterizing the initial orientation of $\uvect{L}_\imag$.
The detector's orientation for both LISA and TianGO is described in Ref.~\cite{Dhurandhar:05}. 

\begin{figure}
  \centering
  \includegraphics[width=0.85\columnwidth]{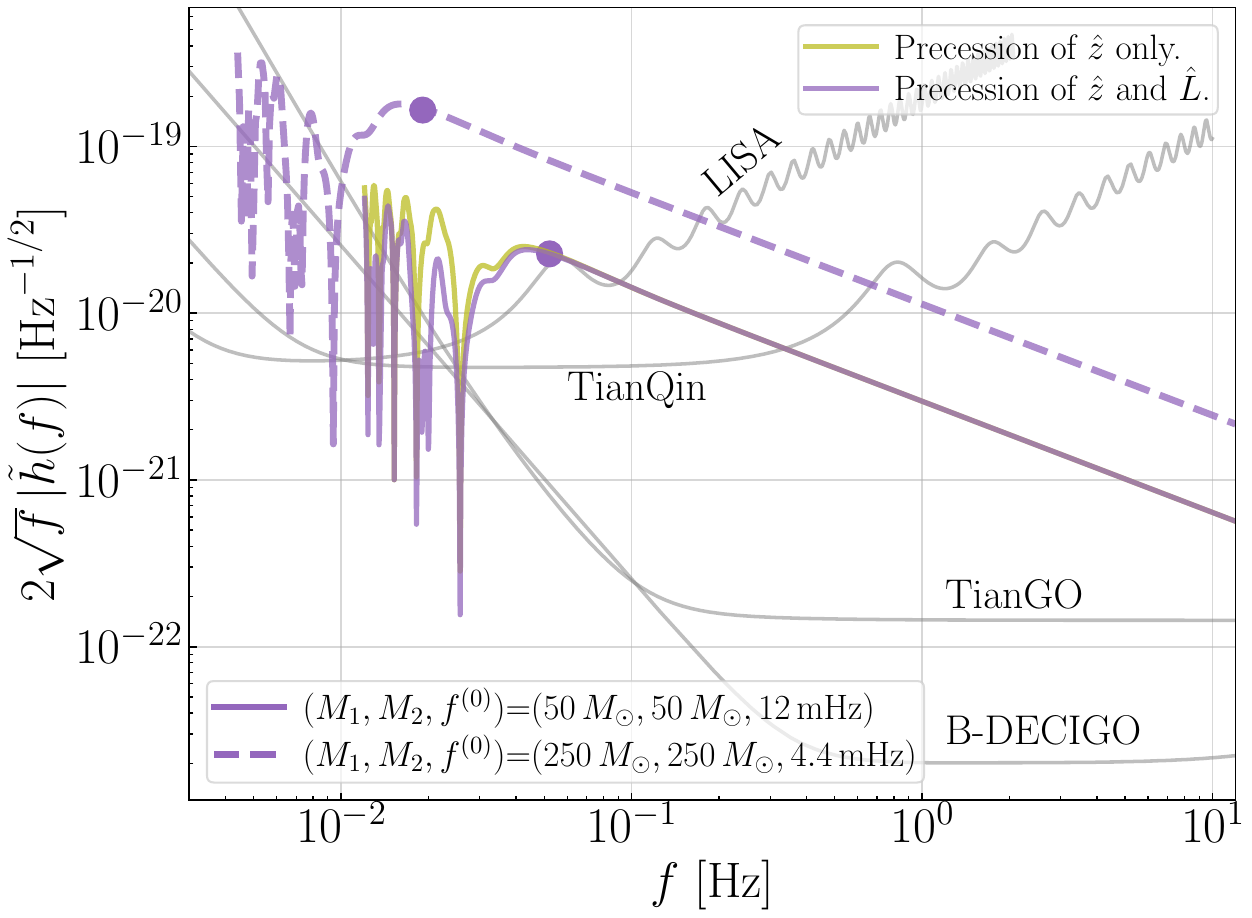}
  \caption{Sample waveforms shown in $2\sqrt{f}|\tilde{h}|$. The olive trace includes variation in the detector's orientation only while the purple ones further include the dS precession. 
  The initial frequency $f^{(0)}$ is chosen such that the binary merges in $T_{\rm obs}=5\,{\rm yr}$ and the dot symbols indicate the instant $0.1\,{\rm yr}$ prior to the merger.
  We assumed $(D_{\rm L}, \overline{\theta}_{\rm S}, \overline{\phi}_{\rm S}, \overline{\theta}_J, \overline{\phi}_J)=(1\,{\rm Gpc}, 33^\circ, 147^\circ, 75^\circ, 150^\circ)$ and $(P_{\rm dS}, \lambda_{\rm L})=(2.7\,{\rm yr}, 45^\circ)$.  
  }
\label{fig:sample_waveform}
\end{figure}

We compare in Fig.~\ref{fig:sample_waveform} sample waveforms with sensitivities of various space-borne detectors. The initial GW frequencies $f^{(0)}$ is chosen such that the inner binary mergers in 5 years, the fiducial value of $T_{\rm obs}$. For a stellar-mass inner binary (solid traces), various missions have similar sensitivities to the precession-induced modulation with the decihertz observatories having a greater total signal-to-noise ratio (SNR). With TianGO's sensitivity, the system corresponding to the purple-solid trace has a total SNR of 80, and an SNR of 13 if we use only the data at least $0.1\,{\rm yr}$ prior to the merger (i.e., integrating from the initial frequency to the dot markers). While the SNR from the final $0.1\,{\rm yr}$ does not directly constrain the precession, it nonetheless reduces the uncertainties on other parameters that are partially degenerate with the precession signatures and is thus critical as well. Similarly, a joint detection of the source with ground-based detectors enhances the sensitivity further. If the inner binary consists of intermediate-mass BHs [the dashed trace; it has a total (early-stage) SNR of 36 (26) in LISA after combining two detectors' responses], then LISA alone would be able to detect the modulations. 

\begin{figure}
  \centering
  \includegraphics[width=0.49\textwidth]{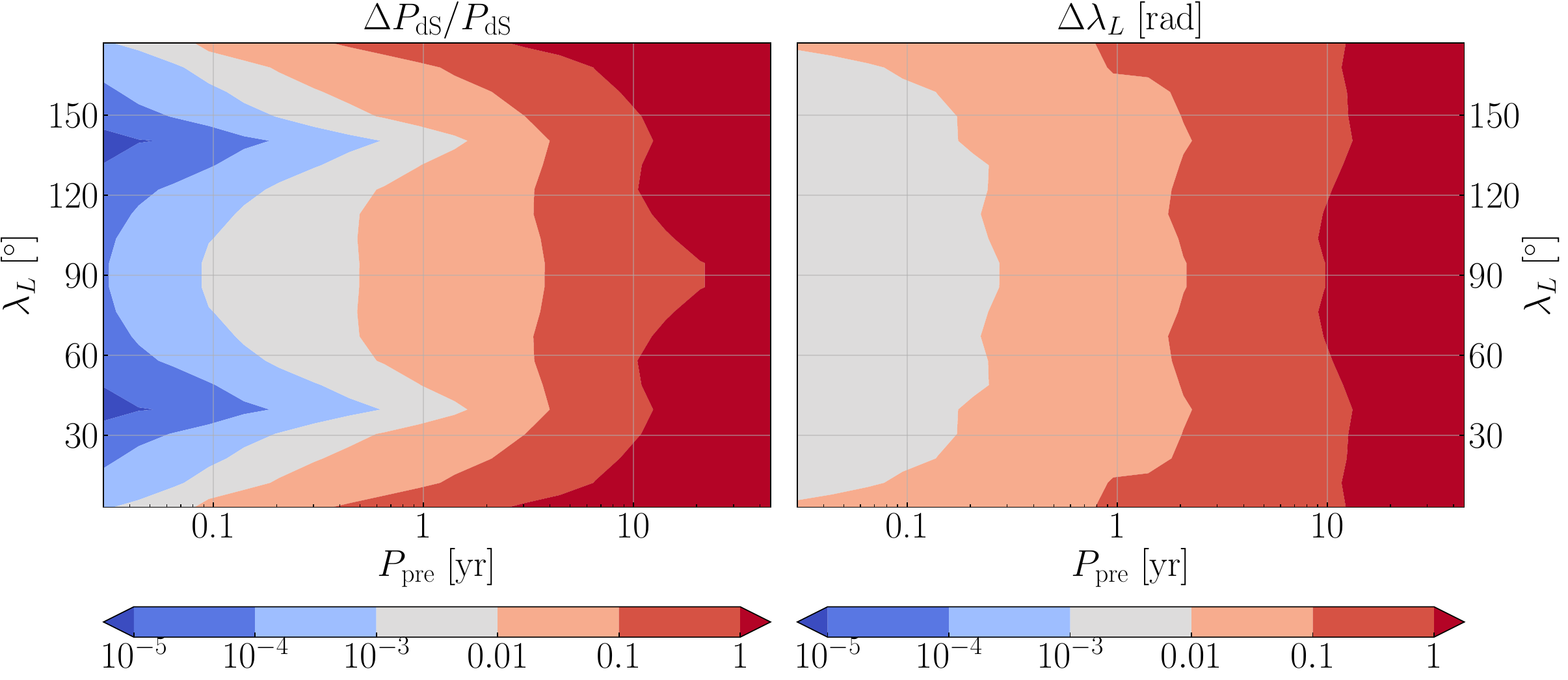}
  \caption{
  PE results assuming a simple-precession problem (no Doppler phase of the outer orbit). We fix the inner binary to have $(M_1, M_2, f^{(0)})=(50\,M_\odot,\, 50\,M_\odot, 12\,{\rm mHz})$. The source's sky location in the solar frame is $(D_L, \overline{\theta}_{\rm S}, \overline{\phi}_{\rm S})=(1\,{\rm Gpc}, 33^\circ, 147^\circ)$ and the orientation of the outer orbit is $(\overline{\theta}_J, \overline{\phi}_J)=(75^\circ, 150^\circ)$. Note the dS precession is detectable if $P_{\rm dS}\lesssim 10\,{\rm yr}$ if the source is detected by TianGO alone.
  }
\label{fig:par_space_simp_prec}
\end{figure}

\begin{figure}
  \centering
  \includegraphics[width=0.49\textwidth]{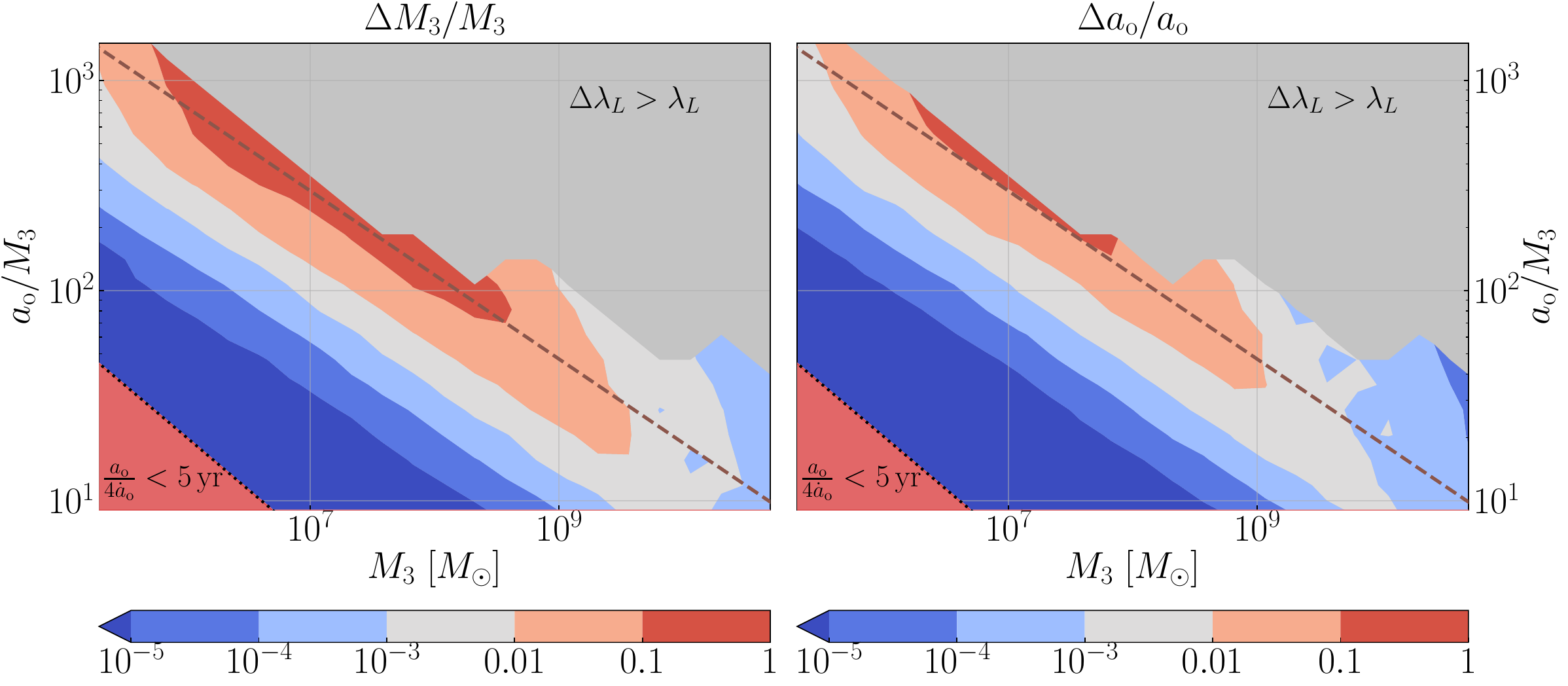}
  \caption{PE results combining both the dS precession and the Doppler phase shift. We assumed an opening angle between the inner and outer orbits of $\lambda_L=45^\circ$ and other parameters are the same as in Fig.~\ref{fig:par_space_simp_prec}. We also plotted the line of $P_{\rm dS}=10\,{\rm yr}$. The grey regions have $\Delta \lambda_L>\lambda_L$ and therefore are excluded. Note $M_3$ can be constrained to $10\%$ if $P_{\rm dS} \simeq 10\,{\rm yr}$ with TianGO alone. The error decreases as $M_3$ increases thanks to the additional information provided by the amplitude of the Doppler phase.}
\label{fig:par_space_full_smbh}
\end{figure}

\emph{Results} -- 
We adopt the Fisher matrix formalism~\cite{Cutler:98} to quantify the detectability.\footnote{As a caveat, we note the Fisher matrix may be inaccurate when the SNR is low~\cite{Vallisneri:08}. Therefore, future studies consider this problem in a full Bayesian framework would be of great value.} We start by considering the parameter-estimation (PE) accuracy of a simple-precession problem (i.e., dropping the Doppler phase due to the outer orbit) and parameterize the modulation in terms of 
$(P_{\rm dS}, \lambda_L, \alpha_0)$.
Our aim is to establish the detectability thresholds for $P_{\rm dS}$ and $\lambda_L$. The results are summarized in Fig.~\ref{fig:par_space_simp_prec} (we have randomized $\alpha_0$ and plotted the median values). Throughout this Section we assume the source is detected by TianGO~\cite{Kuns:19} alone. 

As expected, the accuracy in both $P_{\rm dS}$ and $\lambda_L$  improves as $P_{\rm dS}$ decreases, and at $P_{\rm dS}\simeq 2T_{\rm obs}=10\,{\rm yr}$ we have approximately $\Delta P_{\rm dS}/P_{\rm dS}<1$ and $\Delta \lambda_L<1\,{\rm rad}$, marking the boundary of detectability.  

Note that at $P_{\rm dS}\gtrsim 3\,{\rm yr}$, the error $\Delta P_{\rm dS}$ is smallest when $\lambda_L\simeq 90^\circ$ as it maximizes the variation in the orientation. At smaller $P_{\rm dS}$, the optimal detectability is achieved at $\lambda_L\simeq 40^\circ\simeq \iota_J$ (and also at $140^\circ$). This is thanks to the Thomas phase $\Phi_{\rm T}$. As shown in Ref.~\cite{Apostolatos:94}, when $\uvect{N}$ is inside the precession cone $(|\uvect{L}_\out\cdot\uvect{L}|<|\uvect{L}_\out\cdot \uvect{N}|)$, each precession cycle the Thomas term contributes approximately $(-2\pi\cos\lambda_L)$ to the phase. When $\uvect{L}_\out\cdot\uvect{L}>|\uvect{L}_\out\cdot \uvect{N}|$, however, the contribution per cycle changes sharply to about $2\pi(-\cos\lambda_L + 1)$ 
\footnote{See sec.~IV of Ref.~\cite{BCV} for a treatment that gives the same waveforms without introducing this apparently discontinuous Thomas Precession phase.}. 
Consequently, when $\lambda_L\simeq \iota_J$ (or $\pi-\iota_J$),  $\Phi_{\rm T}$  can be determined with high accuracy. Since the total $\Phi_{\rm T}$ is proportional to the total number of precession cycles, it thus leads to good constraints on $P_{\rm dS}$. 

As we know the detector's orbit, we do not see it significantly interfering with the results when $P_{\rm dS}\simeq 1\,{\rm yr}$. Moreover, the Thomas phase is associated with the precession of $\uvect{L}_\imag$ only~\cite{Cutler:98}, further breaking the potential degeneracy between a changing $\uvect{L}$ and a changing $\uvect{z}$. It is nonetheless crucial to include the detector's motion to constrain $\uvect{N}$~\cite{Cutler:98, Kuns:19}.

We now combine the dS precession with the Doppler shift to study the constraints on the SMBH properties. We use $M_3$ and $a_\out$ as free parameters and write $\Omega_\out$ and $P_{\rm dS}$ in terms of $M_3$ and $a_\out$. The initial phase $\phi^{(0)}$ is included and randomized over. 

The result is shown in Fig.~\ref{fig:par_space_full_smbh}. We only include regions where $\Delta \lambda_L \leq \lambda_L$ so that the signature of precession is unambiguously detected. Note the boundary of $\Delta \lambda_L=\lambda_L$ is broadly consistent with the line of $P_{\rm dS}=10\,{\rm yr}$, agreeing with the results we obtained in the simple-precession analysis. Along the line of $P_{\rm dS}=10\,{\rm yr}$, the fractional error in the SMBH mass is constraned to $\Delta M_3/M_3\sim 10\%$, demonstrating a direct determination of the SMBH property is indeed possible. We further find that $\Delta \log a_\out \simeq \Delta \log M_3/3$ for most of the parameter spaces because $\Omega_\out$ is determined with the highest accuracy among all the parameters describing the modulations. 

Along the line of constant $P_{\rm dS}$, the error decreases with increasing $M_3$. This is because the ``modulation depth'' on the Doppler phase [Eq.~(\ref{eq:Phi_D})] increases with $M_3$. With the Doppler shift alone, we cannot utilize the modulation depth 
due to the unknown $\sin\iota_J$. Once the precession is included, however, $\uvect{L}_\out$ serves as the precession axis of $\uvect{L}_\imag$, allowing the outer orbit's inclination to be inferred. Once we know $\sin\iota_J$, the modulation depth provides another measurement of $a_\out$, enhancing the sensitivity further.  

\emph{Summary and Discussion.} --
Our analysis so far considered detections by TianGO alone. As ground-based detectors are more sensitive to stellar-mass BHs~\cite{Kuns:19}, they could constrain intrinsic parameters with much higher accuracy. We thus estimate the joint-detection effect by still computing the Fisher matrix using a space-borne detector's sensitivity but treating $(\mathcal{M}, \phi_c, t_{c})$ as known parameters. For a system with $(M_3,a_\out)=(10^8\,M_\odot, 100\,M_3)$ and the rest the same as in Fig.~\ref{fig:par_space_full_smbh}, the errors in $(M_3, a_\out)$ can be dramatically improved to $\Delta \log M_3 = 1.7\times10^{-4} (5.2\times 10^{-2})\simeq 3\log a_\out$ assuming the sensitivity of TianGO (LISA). The uncertainty in $\lambda_L$ is also reduced by about a factor of 6 to $\Delta \lambda_L=0.02\,{\rm rad}$ for both TianGO and LISA. If a source instead has $a_\out=300 M_3$ with $P_{\rm pre}\simeq 100\,{\rm yr}$, we find a median error $\Delta \lambda_L = 0.72\,{\rm rad} < \lambda_L$ with LISA's sensitivity after randomizing initial phases, indicating the precession would still be detectable. Knowing the source's distance and sky location further improves the accuracy in $\lambda_L$ by a factor of a few. For $a_\out=300\,M_3$ and LISA's sensitivity, we find $\Delta \lambda_L = 0.16\,{\rm rad}$ in this case.

We assumed both circular inner and outer orbits. In reality, finite eccentricities are expected especially if the inner binary is formed via dynamical channels. One plausible scenario is that both $e_\imag$ and $e_\out$ follow a thermal distribution, with $e_{\imag, \out}^2$ uniform in $[0, 1)$~\cite{Heggie:75, Antonini:12}. 

An elliptic outer orbit enhances the detectability. Note $e_\out$ does not affect the inference accuracy of $\Omega_\out$ and itself can be well constrained from the Doppler shift (as demonstrated in the Supplemental Material). Although the instantaneous precession rate~\cite{Barker:75} should be used for waveform modeling, the secular version [Eq.~(\ref{eq:dS_prec})] nonetheless indicates the qualitative effect of $e_\out$, which is to make the rate greater by a factor of $1/(1-e_\out^2)$. Thus at a fixed $a_\out$ the waveform is modulated by more precession cycles, making its signature more prominent. It also allows a system at greater $a_\out$ to potentially experience a significant modulation (lower panel of Fig.~\ref{fig:periods}).

The eccentricity of the inner orbit $e_\imag$ modifies only the carrier $\tilde{h}_{\rm c}$. Therefore, it affects the results mostly through affecting the overall SNR. Following Ref.~\cite{Barack:04}, for mild eccentricities ($e_\imag\leq 0.7$ at $a_\imag=1.4\times10^{-3}\,{\rm AU}$), we find both the total SNR and that from the early stage ($\geq 0.1\,{\rm yr}$ prior to merger) in fact increase for TianGO, and decreases by a small amount (factor of 3) for LISA.  A more extreme eccentricity would make inner orbit decay too quickly if we fix the initial $a_\imag^{(0)}$. Nonetheless, such a system can merge within $T_{\rm obs}$ starting at much greater initial separations of $\mathcal{O}(0.1)\,{\rm AU}$. From the evolution from $0.1\,{\rm AU}$ to $10^{-3}\,{\rm AU}$ we can still obtain an integration time of more than a year and an SNR of about 5 (with the sensitivity of TianGO). Therefore, our results should not change qualitatively by the inner eccentricity (detailed calculations presented in the Supplemental Material).

We did not include the precessions of $\uvect{L}_\imag$ due to the spins of $M_{1(2)}$. Nevertheless, this should be well distinguishable from the precessions around $\uvect{L}_\out$ thanks to the separation in scales. The spin-induced opening angle is $\lesssim M_1^2/L_\imag\sim 1^\circ$ when $f\sim 0.01\,{\rm Hz}$, in general much smaller than $\Lambda_L$ which distributes approximately uniformly between $0^\circ$ and $180^\circ$ (e.g., Ref.~\cite{Tagawa:20}). Moreover, the spin-induced precession rate is $\sim L_\imag/a_\imag^3$~\cite{Apostolatos:94}, corresponding to a period of $10\,{\rm days}$ when $f=0.01\,{\rm Hz}$, and the period decreases further as the inner binary decays. In contrast, the dS precession around $L_\out$ has a constant and much longer period.

Whereas we used the quadrupole formula for the carrier, our formalism can be readily extended to incorporated more complicated dynamics of the inner binary (higher-order relativistic corrections as well as environmental effects due to gas~\cite{Chen:20} and/or gravitational lensing~\cite{Kocsis:13, Chen:19} that alter the observed chirp mass~\cite{Chen:20b}) by replacing the carrier part with the appropriate $\tilde{h}_{\rm C}(f)$. Similar to the inner eccentricity, changing the carrier affects the detectability of extrinsic modulations mostly through changing the overall SNR. 

To conclude, we demonstrated that the dS precession of $\uvect{L}_\imag$ around $\uvect{L}_\out$ is detectable. The detectability threshold is $P_{\rm dS}\simeq 10\,{\rm yr}$ with space-borne detectors alone and $P_{\rm dS}\simeq 100\,{\rm yr}$ if the source is jointly detected by ground-based detectors. This effect allows a direct determination of the SMBH mass to better than $10\%$ at Gpc distances and applies to both active and quiescent SMBHs. It also constrains the dynamics in galactic nuclei by pinpointing the outer orbit. Future studies incorporating the orbital eccentricities and sub-leading effects, as well as extending the PE to a more rigorous Bayesian framework would be of great value.

\begin{acknowledgments}
\emph{Acknowledgments. }
We thank the helpful comments from Imre Bartos, Karan Jani, and the referees during the preparation of this work.
We are grateful to Hiroyuki Nakano for kindly providing us the sensitivity curve of B-DECIGO. 
H.Y. acknowledges the support by the Sherman Fairchild Foundation. 
Y.C.\ is supported by the Simons Foundation (Award Number 568762), and the National Science Foundation, through Grants PHY-2011961, PHY-2011968, and PHY-1836809. The authors also gratefully acknowledge the computational resources provided by the LIGO Laboratory and supported by NSF grants PHY-0757058 and PHY-0823459.  
\end{acknowledgments}

\bibliography{ref}

\begin{thebibliography}{72}%
\makeatletter
\providecommand \@ifxundefined [1]{%
 \@ifx{#1\undefined}
}%
\providecommand \@ifnum [1]{%
 \ifnum #1\expandafter \@firstoftwo
 \else \expandafter \@secondoftwo
 \fi
}%
\providecommand \@ifx [1]{%
 \ifx #1\expandafter \@firstoftwo
 \else \expandafter \@secondoftwo
 \fi
}%
\providecommand \natexlab [1]{#1}%
\providecommand \enquote  [1]{``#1''}%
\providecommand \bibnamefont  [1]{#1}%
\providecommand \bibfnamefont [1]{#1}%
\providecommand \citenamefont [1]{#1}%
\providecommand \href@noop [0]{\@secondoftwo}%
\providecommand \href [0]{\begingroup \@sanitize@url \@href}%
\providecommand \@href[1]{\@@startlink{#1}\@@href}%
\providecommand \@@href[1]{\endgroup#1\@@endlink}%
\providecommand \@sanitize@url [0]{\catcode `\\12\catcode `\$12\catcode
  `\&12\catcode `\#12\catcode `\^12\catcode `\_12\catcode `\%12\relax}%
\providecommand \@@startlink[1]{}%
\providecommand \@@endlink[0]{}%
\providecommand \url  [0]{\begingroup\@sanitize@url \@url }%
\providecommand \@url [1]{\endgroup\@href {#1}{\urlprefix }}%
\providecommand \urlprefix  [0]{URL }%
\providecommand \Eprint [0]{\href }%
\providecommand \doibase [0]{https://doi.org/}%
\providecommand \selectlanguage [0]{\@gobble}%
\providecommand \bibinfo  [0]{\@secondoftwo}%
\providecommand \bibfield  [0]{\@secondoftwo}%
\providecommand \translation [1]{[#1]}%
\providecommand \BibitemOpen [0]{}%
\providecommand \bibitemStop [0]{}%
\providecommand \bibitemNoStop [0]{.\EOS\space}%
\providecommand \EOS [0]{\spacefactor3000\relax}%
\providecommand \BibitemShut  [1]{\csname bibitem#1\endcsname}%
\let\auto@bib@innerbib\@empty
\bibitem [{\citenamefont {{LIGO Scientific Collaboration}}(2015)}]{LSC:15}%
  \BibitemOpen
  \bibfield  {author} {\bibinfo {author} {\bibnamefont {{LIGO Scientific
  Collaboration}}},\ }\bibfield  {title} {\bibinfo {title} {{Advanced LIGO}},\
  }\href {https://doi.org/10.1088/0264-9381/32/7/074001} {\bibfield  {journal}
  {\bibinfo  {journal} {Classical and Quantum Gravity}\ }\textbf {\bibinfo
  {volume} {32}},\ \bibinfo {eid} {074001} (\bibinfo {year} {2015})},\ \Eprint
  {https://arxiv.org/abs/1411.4547} {arXiv:1411.4547 [gr-qc]} \BibitemShut
  {NoStop}%
\bibitem [{\citenamefont {{Acernese}}\ \emph {et~al.}(2015)\citenamefont
  {{Acernese}}, \citenamefont {{Agathos}}, \citenamefont {{Agatsuma}},
  \citenamefont {{Aisa}}, \citenamefont {{Allemandou}}, \citenamefont
  {{Allocca}}, \citenamefont {{Amarni}}, \citenamefont {{Astone}},\ and\
  \citenamefont {et~al.}}]{Acernese:15}%
  \BibitemOpen
  \bibfield  {author} {\bibinfo {author} {\bibfnamefont {F.}~\bibnamefont
  {{Acernese}}}, \bibinfo {author} {\bibfnamefont {M.}~\bibnamefont
  {{Agathos}}}, \bibinfo {author} {\bibfnamefont {K.}~\bibnamefont
  {{Agatsuma}}}, \bibinfo {author} {\bibfnamefont {D.}~\bibnamefont {{Aisa}}},
  \bibinfo {author} {\bibfnamefont {N.}~\bibnamefont {{Allemandou}}}, \bibinfo
  {author} {\bibfnamefont {A.}~\bibnamefont {{Allocca}}}, \bibinfo {author}
  {\bibfnamefont {J.}~\bibnamefont {{Amarni}}}, \bibinfo {author}
  {\bibfnamefont {P.}~\bibnamefont {{Astone}}},\ and\ \bibinfo {author}
  {\bibnamefont {et~al.}},\ }\bibfield  {title} {\bibinfo {title} {{Advanced
  Virgo: a second-generation interferometric gravitational wave detector}},\
  }\href {https://doi.org/10.1088/0264-9381/32/2/024001} {\bibfield  {journal}
  {\bibinfo  {journal} {Classical and Quantum Gravity}\ }\textbf {\bibinfo
  {volume} {32}},\ \bibinfo {eid} {024001} (\bibinfo {year} {2015})},\ \Eprint
  {https://arxiv.org/abs/1408.3978} {arXiv:1408.3978 [gr-qc]} \BibitemShut
  {NoStop}%
\bibitem [{\citenamefont {{O'Leary}}\ \emph {et~al.}(2009)\citenamefont
  {{O'Leary}}, \citenamefont {{Kocsis}},\ and\ \citenamefont
  {{Loeb}}}]{OLeary:09}%
  \BibitemOpen
  \bibfield  {author} {\bibinfo {author} {\bibfnamefont {R.~M.}\ \bibnamefont
  {{O'Leary}}}, \bibinfo {author} {\bibfnamefont {B.}~\bibnamefont
  {{Kocsis}}},\ and\ \bibinfo {author} {\bibfnamefont {A.}~\bibnamefont
  {{Loeb}}},\ }\bibfield  {title} {\bibinfo {title} {{Gravitational waves from
  scattering of stellar-mass black holes in galactic nuclei}},\ }\href
  {https://doi.org/10.1111/j.1365-2966.2009.14653.x} {\bibfield  {journal}
  {\bibinfo  {journal} {\mnras}\ }\textbf {\bibinfo {volume} {395}},\ \bibinfo
  {pages} {2127} (\bibinfo {year} {2009})},\ \Eprint
  {https://arxiv.org/abs/0807.2638} {arXiv:0807.2638 [astro-ph]} \BibitemShut
  {NoStop}%
\bibitem [{\citenamefont {{Antonini}}\ and\ \citenamefont
  {{Perets}}(2012)}]{Antonini:12}%
  \BibitemOpen
  \bibfield  {author} {\bibinfo {author} {\bibfnamefont {F.}~\bibnamefont
  {{Antonini}}}\ and\ \bibinfo {author} {\bibfnamefont {H.~B.}\ \bibnamefont
  {{Perets}}},\ }\bibfield  {title} {\bibinfo {title} {{Secular Evolution of
  Compact Binaries near Massive Black Holes: Gravitational Wave Sources and
  Other Exotica}},\ }\href {https://doi.org/10.1088/0004-637X/757/1/27}
  {\bibfield  {journal} {\bibinfo  {journal} {\apj}\ }\textbf {\bibinfo
  {volume} {757}},\ \bibinfo {eid} {27} (\bibinfo {year} {2012})},\ \Eprint
  {https://arxiv.org/abs/1203.2938} {arXiv:1203.2938 [astro-ph.GA]}
  \BibitemShut {NoStop}%
\bibitem [{\citenamefont {{Antonini}}\ and\ \citenamefont
  {{Rasio}}(2016)}]{Antonini:16}%
  \BibitemOpen
  \bibfield  {author} {\bibinfo {author} {\bibfnamefont {F.}~\bibnamefont
  {{Antonini}}}\ and\ \bibinfo {author} {\bibfnamefont {F.~A.}\ \bibnamefont
  {{Rasio}}},\ }\bibfield  {title} {\bibinfo {title} {{Merging Black Hole
  Binaries in Galactic Nuclei: Implications for Advanced-LIGO Detections}},\
  }\href {https://doi.org/10.3847/0004-637X/831/2/187} {\bibfield  {journal}
  {\bibinfo  {journal} {\apj}\ }\textbf {\bibinfo {volume} {831}},\ \bibinfo
  {eid} {187} (\bibinfo {year} {2016})},\ \Eprint
  {https://arxiv.org/abs/1606.04889} {arXiv:1606.04889 [astro-ph.HE]}
  \BibitemShut {NoStop}%
\bibitem [{\citenamefont {{Petrovich}}\ and\ \citenamefont
  {{Antonini}}(2017)}]{Petrovich:17}%
  \BibitemOpen
  \bibfield  {author} {\bibinfo {author} {\bibfnamefont {C.}~\bibnamefont
  {{Petrovich}}}\ and\ \bibinfo {author} {\bibfnamefont {F.}~\bibnamefont
  {{Antonini}}},\ }\bibfield  {title} {\bibinfo {title} {{Greatly Enhanced
  Merger Rates of Compact-object Binaries in Non-spherical Nuclear Star
  Clusters}},\ }\href {https://doi.org/10.3847/1538-4357/aa8628} {\bibfield
  {journal} {\bibinfo  {journal} {\apj}\ }\textbf {\bibinfo {volume} {846}},\
  \bibinfo {eid} {146} (\bibinfo {year} {2017})},\ \Eprint
  {https://arxiv.org/abs/1705.05848} {arXiv:1705.05848 [astro-ph.HE]}
  \BibitemShut {NoStop}%
\bibitem [{\citenamefont {{Leigh}}\ \emph {et~al.}(2018)\citenamefont
  {{Leigh}}, \citenamefont {{Geller}}, \citenamefont {{McKernan}},
  \citenamefont {{Ford}}, \citenamefont {{Mac Low}}, \citenamefont
  {{Bellovary}}, \citenamefont {{Haiman}}, \citenamefont {{Lyra}},
  \citenamefont {{Samsing}}, \citenamefont {{O'Dowd}}, \citenamefont
  {{Kocsis}},\ and\ \citenamefont {{Endlich}}}]{Leigh:18}%
  \BibitemOpen
  \bibfield  {author} {\bibinfo {author} {\bibfnamefont {N.~W.~C.}\
  \bibnamefont {{Leigh}}}, \bibinfo {author} {\bibfnamefont {A.~M.}\
  \bibnamefont {{Geller}}}, \bibinfo {author} {\bibfnamefont {B.}~\bibnamefont
  {{McKernan}}}, \bibinfo {author} {\bibfnamefont {K.~E.~S.}\ \bibnamefont
  {{Ford}}}, \bibinfo {author} {\bibfnamefont {M.~M.}\ \bibnamefont {{Mac
  Low}}}, \bibinfo {author} {\bibfnamefont {J.}~\bibnamefont {{Bellovary}}},
  \bibinfo {author} {\bibfnamefont {Z.}~\bibnamefont {{Haiman}}}, \bibinfo
  {author} {\bibfnamefont {W.}~\bibnamefont {{Lyra}}}, \bibinfo {author}
  {\bibfnamefont {J.}~\bibnamefont {{Samsing}}}, \bibinfo {author}
  {\bibfnamefont {M.}~\bibnamefont {{O'Dowd}}}, \bibinfo {author}
  {\bibfnamefont {B.}~\bibnamefont {{Kocsis}}},\ and\ \bibinfo {author}
  {\bibfnamefont {S.}~\bibnamefont {{Endlich}}},\ }\bibfield  {title} {\bibinfo
  {title} {{On the rate of black hole binary mergers in galactic nuclei due to
  dynamical hardening}},\ }\href {https://doi.org/10.1093/mnras/stx3134}
  {\bibfield  {journal} {\bibinfo  {journal} {\mnras}\ }\textbf {\bibinfo
  {volume} {474}},\ \bibinfo {pages} {5672} (\bibinfo {year} {2018})},\ \Eprint
  {https://arxiv.org/abs/1711.10494} {arXiv:1711.10494 [astro-ph.GA]}
  \BibitemShut {NoStop}%
\bibitem [{\citenamefont {{Chen}}\ and\ \citenamefont {{Han}}(2018)}]{Chen:18}%
  \BibitemOpen
  \bibfield  {author} {\bibinfo {author} {\bibfnamefont {X.}~\bibnamefont
  {{Chen}}}\ and\ \bibinfo {author} {\bibfnamefont {W.-B.}\ \bibnamefont
  {{Han}}},\ }\bibfield  {title} {\bibinfo {title} {{Extreme-mass-ratio
  inspirals produced by tidal capture of binary black holes}},\ }\href
  {https://doi.org/10.1038/s42005-018-0053-0} {\bibfield  {journal} {\bibinfo
  {journal} {Communications Physics}\ }\textbf {\bibinfo {volume} {1}},\
  \bibinfo {eid} {53} (\bibinfo {year} {2018})},\ \Eprint
  {https://arxiv.org/abs/1801.05780} {arXiv:1801.05780 [astro-ph.HE]}
  \BibitemShut {NoStop}%
\bibitem [{\citenamefont {{Fragione}}\ \emph {et~al.}(2019)\citenamefont
  {{Fragione}}, \citenamefont {{Leigh}},\ and\ \citenamefont
  {{Perna}}}]{Fragione:19}%
  \BibitemOpen
  \bibfield  {author} {\bibinfo {author} {\bibfnamefont {G.}~\bibnamefont
  {{Fragione}}}, \bibinfo {author} {\bibfnamefont {N.~W.~C.}\ \bibnamefont
  {{Leigh}}},\ and\ \bibinfo {author} {\bibfnamefont {R.}~\bibnamefont
  {{Perna}}},\ }\bibfield  {title} {\bibinfo {title} {{Black hole and neutron
  star mergers in galactic nuclei: the role of triples}},\ }\href
  {https://doi.org/10.1093/mnras/stz1803} {\bibfield  {journal} {\bibinfo
  {journal} {\mnras}\ }\textbf {\bibinfo {volume} {488}},\ \bibinfo {pages}
  {2825} (\bibinfo {year} {2019})},\ \Eprint {https://arxiv.org/abs/1903.09160}
  {arXiv:1903.09160 [astro-ph.GA]} \BibitemShut {NoStop}%
\bibitem [{\citenamefont {{McKernan}}\ \emph {et~al.}(2012)\citenamefont
  {{McKernan}}, \citenamefont {{Ford}}, \citenamefont {{Lyra}},\ and\
  \citenamefont {{Perets}}}]{McKernan:12}%
  \BibitemOpen
  \bibfield  {author} {\bibinfo {author} {\bibfnamefont {B.}~\bibnamefont
  {{McKernan}}}, \bibinfo {author} {\bibfnamefont {K.~E.~S.}\ \bibnamefont
  {{Ford}}}, \bibinfo {author} {\bibfnamefont {W.}~\bibnamefont {{Lyra}}},\
  and\ \bibinfo {author} {\bibfnamefont {H.~B.}\ \bibnamefont {{Perets}}},\
  }\bibfield  {title} {\bibinfo {title} {{Intermediate mass black holes in AGN
  discs - I. Production and growth}},\ }\href
  {https://doi.org/10.1111/j.1365-2966.2012.21486.x} {\bibfield  {journal}
  {\bibinfo  {journal} {\mnras}\ }\textbf {\bibinfo {volume} {425}},\ \bibinfo
  {pages} {460} (\bibinfo {year} {2012})},\ \Eprint
  {https://arxiv.org/abs/1206.2309} {arXiv:1206.2309 [astro-ph.GA]}
  \BibitemShut {NoStop}%
\bibitem [{\citenamefont {{Bartos}}\ \emph {et~al.}(2017)\citenamefont
  {{Bartos}}, \citenamefont {{Kocsis}}, \citenamefont {{Haiman}},\ and\
  \citenamefont {{M{\'a}rka}}}]{Bartos:17}%
  \BibitemOpen
  \bibfield  {author} {\bibinfo {author} {\bibfnamefont {I.}~\bibnamefont
  {{Bartos}}}, \bibinfo {author} {\bibfnamefont {B.}~\bibnamefont {{Kocsis}}},
  \bibinfo {author} {\bibfnamefont {Z.}~\bibnamefont {{Haiman}}},\ and\
  \bibinfo {author} {\bibfnamefont {S.}~\bibnamefont {{M{\'a}rka}}},\
  }\bibfield  {title} {\bibinfo {title} {{Rapid and Bright Stellar-mass Binary
  Black Hole Mergers in Active Galactic Nuclei}},\ }\href
  {https://doi.org/10.3847/1538-4357/835/2/165} {\bibfield  {journal} {\bibinfo
   {journal} {\apj}\ }\textbf {\bibinfo {volume} {835}},\ \bibinfo {eid} {165}
  (\bibinfo {year} {2017})},\ \Eprint {https://arxiv.org/abs/1602.03831}
  {arXiv:1602.03831 [astro-ph.HE]} \BibitemShut {NoStop}%
\bibitem [{\citenamefont {{Stone}}\ \emph {et~al.}(2017)\citenamefont
  {{Stone}}, \citenamefont {{Metzger}},\ and\ \citenamefont
  {{Haiman}}}]{Stone:17}%
  \BibitemOpen
  \bibfield  {author} {\bibinfo {author} {\bibfnamefont {N.~C.}\ \bibnamefont
  {{Stone}}}, \bibinfo {author} {\bibfnamefont {B.~D.}\ \bibnamefont
  {{Metzger}}},\ and\ \bibinfo {author} {\bibfnamefont {Z.}~\bibnamefont
  {{Haiman}}},\ }\bibfield  {title} {\bibinfo {title} {{Assisted inspirals of
  stellar mass black holes embedded in AGN discs: solving the `final au
  problem'}},\ }\href {https://doi.org/10.1093/mnras/stw2260} {\bibfield
  {journal} {\bibinfo  {journal} {\mnras}\ }\textbf {\bibinfo {volume} {464}},\
  \bibinfo {pages} {946} (\bibinfo {year} {2017})},\ \Eprint
  {https://arxiv.org/abs/1602.04226} {arXiv:1602.04226 [astro-ph.GA]}
  \BibitemShut {NoStop}%
\bibitem [{\citenamefont {{McKernan}}\ \emph {et~al.}(2018)\citenamefont
  {{McKernan}}, \citenamefont {{Ford}}, \citenamefont {{Bellovary}},
  \citenamefont {{Leigh}}, \citenamefont {{Haiman}}, \citenamefont {{Kocsis}},
  \citenamefont {{Lyra}}, \citenamefont {{Mac Low}}, \citenamefont {{Metzger}},
  \citenamefont {{O'Dowd}}, \citenamefont {{Endlich}},\ and\ \citenamefont
  {{Rosen}}}]{McKernan:18}%
  \BibitemOpen
  \bibfield  {author} {\bibinfo {author} {\bibfnamefont {B.}~\bibnamefont
  {{McKernan}}}, \bibinfo {author} {\bibfnamefont {K.~E.~S.}\ \bibnamefont
  {{Ford}}}, \bibinfo {author} {\bibfnamefont {J.}~\bibnamefont {{Bellovary}}},
  \bibinfo {author} {\bibfnamefont {N.~W.~C.}\ \bibnamefont {{Leigh}}},
  \bibinfo {author} {\bibfnamefont {Z.}~\bibnamefont {{Haiman}}}, \bibinfo
  {author} {\bibfnamefont {B.}~\bibnamefont {{Kocsis}}}, \bibinfo {author}
  {\bibfnamefont {W.}~\bibnamefont {{Lyra}}}, \bibinfo {author} {\bibfnamefont
  {M.~M.}\ \bibnamefont {{Mac Low}}}, \bibinfo {author} {\bibfnamefont
  {B.}~\bibnamefont {{Metzger}}}, \bibinfo {author} {\bibfnamefont
  {M.}~\bibnamefont {{O'Dowd}}}, \bibinfo {author} {\bibfnamefont
  {S.}~\bibnamefont {{Endlich}}},\ and\ \bibinfo {author} {\bibfnamefont
  {D.~J.}\ \bibnamefont {{Rosen}}},\ }\bibfield  {title} {\bibinfo {title}
  {{Constraining Stellar-mass Black Hole Mergers in AGN Disks Detectable with
  LIGO}},\ }\href {https://doi.org/10.3847/1538-4357/aadae5} {\bibfield
  {journal} {\bibinfo  {journal} {\apj}\ }\textbf {\bibinfo {volume} {866}},\
  \bibinfo {eid} {66} (\bibinfo {year} {2018})},\ \Eprint
  {https://arxiv.org/abs/1702.07818} {arXiv:1702.07818 [astro-ph.HE]}
  \BibitemShut {NoStop}%
\bibitem [{\citenamefont {{Tagawa}}\ \emph {et~al.}(2019)\citenamefont
  {{Tagawa}}, \citenamefont {{Haiman}},\ and\ \citenamefont
  {{Kocsis}}}]{Tagawa:19}%
  \BibitemOpen
  \bibfield  {author} {\bibinfo {author} {\bibfnamefont {H.}~\bibnamefont
  {{Tagawa}}}, \bibinfo {author} {\bibfnamefont {Z.}~\bibnamefont {{Haiman}}},\
  and\ \bibinfo {author} {\bibfnamefont {B.}~\bibnamefont {{Kocsis}}},\
  }\bibfield  {title} {\bibinfo {title} {{Formation and Evolution of Compact
  Object Binaries in AGN Disks}},\ }\href@noop {} {\bibfield  {journal}
  {\bibinfo  {journal} {arXiv e-prints}\ ,\ \bibinfo {eid} {arXiv:1912.08218}}
  (\bibinfo {year} {2019})},\ \Eprint {https://arxiv.org/abs/1912.08218}
  {arXiv:1912.08218 [astro-ph.GA]} \BibitemShut {NoStop}%
\bibitem [{\citenamefont {{McKernan}}\ \emph {et~al.}(2019)\citenamefont
  {{McKernan}}, \citenamefont {{Ford}}, \citenamefont {{Bartos}}, \citenamefont
  {{Graham}}, \citenamefont {{Lyra}}, \citenamefont {{Marka}}, \citenamefont
  {{Marka}}, \citenamefont {{Ross}}, \citenamefont {{Stern}},\ and\
  \citenamefont {{Yang}}}]{McKernan:19}%
  \BibitemOpen
  \bibfield  {author} {\bibinfo {author} {\bibfnamefont {B.}~\bibnamefont
  {{McKernan}}}, \bibinfo {author} {\bibfnamefont {K.~E.~S.}\ \bibnamefont
  {{Ford}}}, \bibinfo {author} {\bibfnamefont {I.}~\bibnamefont {{Bartos}}},
  \bibinfo {author} {\bibfnamefont {M.~J.}\ \bibnamefont {{Graham}}}, \bibinfo
  {author} {\bibfnamefont {W.}~\bibnamefont {{Lyra}}}, \bibinfo {author}
  {\bibfnamefont {S.}~\bibnamefont {{Marka}}}, \bibinfo {author} {\bibfnamefont
  {Z.}~\bibnamefont {{Marka}}}, \bibinfo {author} {\bibfnamefont {N.~P.}\
  \bibnamefont {{Ross}}}, \bibinfo {author} {\bibfnamefont {D.}~\bibnamefont
  {{Stern}}},\ and\ \bibinfo {author} {\bibfnamefont {Y.}~\bibnamefont
  {{Yang}}},\ }\bibfield  {title} {\bibinfo {title} {{Ram-pressure Stripping of
  a Kicked Hill Sphere: Prompt Electromagnetic Emission from the Merger of
  Stellar Mass Black Holes in an AGN Accretion Disk}},\ }\href
  {https://doi.org/10.3847/2041-8213/ab4886} {\bibfield  {journal} {\bibinfo
  {journal} {\apjl}\ }\textbf {\bibinfo {volume} {884}},\ \bibinfo {eid} {L50}
  (\bibinfo {year} {2019})},\ \Eprint {https://arxiv.org/abs/1907.03746}
  {arXiv:1907.03746 [astro-ph.HE]} \BibitemShut {NoStop}%
\bibitem [{\citenamefont {{Yang}}\ \emph {et~al.}(2019)\citenamefont {{Yang}},
  \citenamefont {{Bartos}}, \citenamefont {{Gayathri}}, \citenamefont {{Ford}},
  \citenamefont {{Haiman}}, \citenamefont {{Klimenko}}, \citenamefont
  {{Kocsis}}, \citenamefont {{M{\'a}rka}}, \citenamefont {{M{\'a}rka}},
  \citenamefont {{McKernan}},\ and\ \citenamefont {{O'Shaughnessy}}}]{Yang:19}%
  \BibitemOpen
  \bibfield  {author} {\bibinfo {author} {\bibfnamefont {Y.}~\bibnamefont
  {{Yang}}}, \bibinfo {author} {\bibfnamefont {I.}~\bibnamefont {{Bartos}}},
  \bibinfo {author} {\bibfnamefont {V.}~\bibnamefont {{Gayathri}}}, \bibinfo
  {author} {\bibfnamefont {K.~E.~S.}\ \bibnamefont {{Ford}}}, \bibinfo {author}
  {\bibfnamefont {Z.}~\bibnamefont {{Haiman}}}, \bibinfo {author}
  {\bibfnamefont {S.}~\bibnamefont {{Klimenko}}}, \bibinfo {author}
  {\bibfnamefont {B.}~\bibnamefont {{Kocsis}}}, \bibinfo {author}
  {\bibfnamefont {S.}~\bibnamefont {{M{\'a}rka}}}, \bibinfo {author}
  {\bibfnamefont {Z.}~\bibnamefont {{M{\'a}rka}}}, \bibinfo {author}
  {\bibfnamefont {B.}~\bibnamefont {{McKernan}}},\ and\ \bibinfo {author}
  {\bibfnamefont {R.}~\bibnamefont {{O'Shaughnessy}}},\ }\bibfield  {title}
  {\bibinfo {title} {{Hierarchical Black Hole Mergers in Active Galactic
  Nuclei}},\ }\href {https://doi.org/10.1103/PhysRevLett.123.181101} {\bibfield
   {journal} {\bibinfo  {journal} {\prl}\ }\textbf {\bibinfo {volume} {123}},\
  \bibinfo {eid} {181101} (\bibinfo {year} {2019})},\ \Eprint
  {https://arxiv.org/abs/1906.09281} {arXiv:1906.09281 [astro-ph.HE]}
  \BibitemShut {NoStop}%
\bibitem [{\citenamefont {{Secunda}}\ \emph {et~al.}(2019)\citenamefont
  {{Secunda}}, \citenamefont {{Bellovary}}, \citenamefont {{Mac Low}},
  \citenamefont {{Ford}}, \citenamefont {{McKernan}}, \citenamefont {{Leigh}},
  \citenamefont {{Lyra}},\ and\ \citenamefont {{S{\'a}ndor}}}]{Secunda:19}%
  \BibitemOpen
  \bibfield  {author} {\bibinfo {author} {\bibfnamefont {A.}~\bibnamefont
  {{Secunda}}}, \bibinfo {author} {\bibfnamefont {J.}~\bibnamefont
  {{Bellovary}}}, \bibinfo {author} {\bibfnamefont {M.-M.}\ \bibnamefont {{Mac
  Low}}}, \bibinfo {author} {\bibfnamefont {K.~E.~S.}\ \bibnamefont {{Ford}}},
  \bibinfo {author} {\bibfnamefont {B.}~\bibnamefont {{McKernan}}}, \bibinfo
  {author} {\bibfnamefont {N.~W.~C.}\ \bibnamefont {{Leigh}}}, \bibinfo
  {author} {\bibfnamefont {W.}~\bibnamefont {{Lyra}}},\ and\ \bibinfo {author}
  {\bibfnamefont {Z.}~\bibnamefont {{S{\'a}ndor}}},\ }\bibfield  {title}
  {\bibinfo {title} {{Orbital Migration of Interacting Stellar Mass Black Holes
  in Disks around Supermassive Black Holes}},\ }\href
  {https://doi.org/10.3847/1538-4357/ab20ca} {\bibfield  {journal} {\bibinfo
  {journal} {\apj}\ }\textbf {\bibinfo {volume} {878}},\ \bibinfo {eid} {85}
  (\bibinfo {year} {2019})},\ \Eprint {https://arxiv.org/abs/1807.02859}
  {arXiv:1807.02859 [astro-ph.HE]} \BibitemShut {NoStop}%
\bibitem [{\citenamefont {{Bellm}}\ \emph {et~al.}(2019)\citenamefont
  {{Bellm}}, \citenamefont {{Kulkarni}}, \citenamefont {{Graham}},
  \citenamefont {{Dekany}}, \citenamefont {{Smith}}, \citenamefont {{Riddle}},
  \citenamefont {{Masci}}, \citenamefont {{Helou}},\ and\ \citenamefont
  {et~al.}}]{Bellm:19}%
  \BibitemOpen
  \bibfield  {author} {\bibinfo {author} {\bibfnamefont {E.~C.}\ \bibnamefont
  {{Bellm}}}, \bibinfo {author} {\bibfnamefont {S.~R.}\ \bibnamefont
  {{Kulkarni}}}, \bibinfo {author} {\bibfnamefont {M.~J.}\ \bibnamefont
  {{Graham}}}, \bibinfo {author} {\bibfnamefont {R.}~\bibnamefont {{Dekany}}},
  \bibinfo {author} {\bibfnamefont {R.~M.}\ \bibnamefont {{Smith}}}, \bibinfo
  {author} {\bibfnamefont {R.}~\bibnamefont {{Riddle}}}, \bibinfo {author}
  {\bibfnamefont {F.~J.}\ \bibnamefont {{Masci}}}, \bibinfo {author}
  {\bibfnamefont {G.}~\bibnamefont {{Helou}}},\ and\ \bibinfo {author}
  {\bibnamefont {et~al.}},\ }\bibfield  {title} {\bibinfo {title} {{The Zwicky
  Transient Facility: System Overview, Performance, and First Results}},\
  }\href {https://doi.org/10.1088/1538-3873/aaecbe} {\bibfield  {journal}
  {\bibinfo  {journal} {\pasp}\ }\textbf {\bibinfo {volume} {131}},\ \bibinfo
  {pages} {018002} (\bibinfo {year} {2019})},\ \Eprint
  {https://arxiv.org/abs/1902.01932} {arXiv:1902.01932 [astro-ph.IM]}
  \BibitemShut {NoStop}%
\bibitem [{\citenamefont {{Graham}}\ \emph {et~al.}(2019)\citenamefont
  {{Graham}}, \citenamefont {{Kulkarni}}, \citenamefont {{Bellm}},
  \citenamefont {{Adams}}, \citenamefont {{Barbarino}}, \citenamefont
  {{Blagorodnova}}, \citenamefont {{Bodewits}}, \citenamefont {{Bolin}},\ and\
  \citenamefont {et~al.}}]{Graham:19}%
  \BibitemOpen
  \bibfield  {author} {\bibinfo {author} {\bibfnamefont {M.~J.}\ \bibnamefont
  {{Graham}}}, \bibinfo {author} {\bibfnamefont {S.~R.}\ \bibnamefont
  {{Kulkarni}}}, \bibinfo {author} {\bibfnamefont {E.~C.}\ \bibnamefont
  {{Bellm}}}, \bibinfo {author} {\bibfnamefont {S.~M.}\ \bibnamefont
  {{Adams}}}, \bibinfo {author} {\bibfnamefont {C.}~\bibnamefont
  {{Barbarino}}}, \bibinfo {author} {\bibfnamefont {N.}~\bibnamefont
  {{Blagorodnova}}}, \bibinfo {author} {\bibfnamefont {D.}~\bibnamefont
  {{Bodewits}}}, \bibinfo {author} {\bibfnamefont {B.}~\bibnamefont
  {{Bolin}}},\ and\ \bibinfo {author} {\bibnamefont {et~al.}},\ }\bibfield
  {title} {\bibinfo {title} {{The Zwicky Transient Facility: Science
  Objectives}},\ }\href {https://doi.org/10.1088/1538-3873/ab006c} {\bibfield
  {journal} {\bibinfo  {journal} {\pasp}\ }\textbf {\bibinfo {volume} {131}},\
  \bibinfo {pages} {078001} (\bibinfo {year} {2019})},\ \Eprint
  {https://arxiv.org/abs/1902.01945} {arXiv:1902.01945 [astro-ph.IM]}
  \BibitemShut {NoStop}%
\bibitem [{\citenamefont {Graham}\ \emph {et~al.}(2020)\citenamefont {Graham},
  \citenamefont {Ford}, \citenamefont {McKernan}, \citenamefont {Ross},
  \citenamefont {Stern}, \citenamefont {Burdge}, \citenamefont {Coughlin},
  \citenamefont {Djorgovski},\ and\ \citenamefont {et~al.}}]{Graham:20}%
  \BibitemOpen
  \bibfield  {author} {\bibinfo {author} {\bibfnamefont {M.~J.}\ \bibnamefont
  {Graham}}, \bibinfo {author} {\bibfnamefont {K.~E.~S.}\ \bibnamefont {Ford}},
  \bibinfo {author} {\bibfnamefont {B.}~\bibnamefont {McKernan}}, \bibinfo
  {author} {\bibfnamefont {N.~P.}\ \bibnamefont {Ross}}, \bibinfo {author}
  {\bibfnamefont {D.}~\bibnamefont {Stern}}, \bibinfo {author} {\bibfnamefont
  {K.}~\bibnamefont {Burdge}}, \bibinfo {author} {\bibfnamefont
  {M.}~\bibnamefont {Coughlin}}, \bibinfo {author} {\bibfnamefont {S.~G.}\
  \bibnamefont {Djorgovski}},\ and\ \bibinfo {author} {\bibnamefont {et~al.}},\
  }\bibfield  {title} {\bibinfo {title} {Candidate electromagnetic counterpart
  to the binary black hole merger gravitational-wave event s190521g},\ }\href
  {https://doi.org/10.1103/PhysRevLett.124.251102} {\bibfield  {journal}
  {\bibinfo  {journal} {Phys. Rev. Lett.}\ }\textbf {\bibinfo {volume} {124}},\
  \bibinfo {pages} {251102} (\bibinfo {year} {2020})}\BibitemShut {NoStop}%
\bibitem [{\citenamefont {Abbott}\ \emph
  {et~al.}(2020{\natexlab{a}})\citenamefont {Abbott}, \citenamefont {Abbott},
  \citenamefont {Abraham}, \citenamefont {Acernese}, \citenamefont {Ackley},
  \citenamefont {Adams}, \citenamefont {Adhikari}, \citenamefont {Adya},\ and\
  \citenamefont {et~al.}}]{GW190521a}%
  \BibitemOpen
  \bibfield  {author} {\bibinfo {author} {\bibfnamefont {R.}~\bibnamefont
  {Abbott}}, \bibinfo {author} {\bibfnamefont {T.~D.}\ \bibnamefont {Abbott}},
  \bibinfo {author} {\bibfnamefont {S.}~\bibnamefont {Abraham}}, \bibinfo
  {author} {\bibfnamefont {F.}~\bibnamefont {Acernese}}, \bibinfo {author}
  {\bibfnamefont {K.}~\bibnamefont {Ackley}}, \bibinfo {author} {\bibfnamefont
  {C.}~\bibnamefont {Adams}}, \bibinfo {author} {\bibfnamefont {R.~X.}\
  \bibnamefont {Adhikari}}, \bibinfo {author} {\bibfnamefont {V.~B.}\
  \bibnamefont {Adya}},\ and\ \bibinfo {author} {\bibnamefont {et~al.}}
  (\bibinfo {collaboration} {LIGO Scientific Collaboration and Virgo
  Collaboration}),\ }\bibfield  {title} {\bibinfo {title} {Gw190521: A binary
  black hole merger with a total mass of $150\text{ }\text{
  }{M}_{\ensuremath{\bigodot}}$},\ }\href
  {https://doi.org/10.1103/PhysRevLett.125.101102} {\bibfield  {journal}
  {\bibinfo  {journal} {Phys. Rev. Lett.}\ }\textbf {\bibinfo {volume} {125}},\
  \bibinfo {pages} {101102} (\bibinfo {year} {2020}{\natexlab{a}})}\BibitemShut
  {NoStop}%
\bibitem [{\citenamefont {Abbott}\ \emph
  {et~al.}(2020{\natexlab{b}})\citenamefont {Abbott}, \citenamefont {Abbott},
  \citenamefont {Abraham}, \citenamefont {Acernese}, \citenamefont {Ackley},
  \citenamefont {Adams}, \citenamefont {Adhikari}, \citenamefont {Adya},\ and\
  \citenamefont {et~al.}}]{GW190521b}%
  \BibitemOpen
  \bibfield  {author} {\bibinfo {author} {\bibfnamefont {R.}~\bibnamefont
  {Abbott}}, \bibinfo {author} {\bibfnamefont {T.~D.}\ \bibnamefont {Abbott}},
  \bibinfo {author} {\bibfnamefont {S.}~\bibnamefont {Abraham}}, \bibinfo
  {author} {\bibfnamefont {F.}~\bibnamefont {Acernese}}, \bibinfo {author}
  {\bibfnamefont {K.}~\bibnamefont {Ackley}}, \bibinfo {author} {\bibfnamefont
  {C.}~\bibnamefont {Adams}}, \bibinfo {author} {\bibfnamefont {R.~X.}\
  \bibnamefont {Adhikari}}, \bibinfo {author} {\bibfnamefont {V.~B.}\
  \bibnamefont {Adya}},\ and\ \bibinfo {author} {\bibnamefont {et~al.}},\
  }\bibfield  {title} {\bibinfo {title} {Properties and astrophysical
  implications of the 150 m $\odot$ binary black hole merger {GW}190521},\
  }\href {https://doi.org/10.3847/2041-8213/aba493} {\bibfield  {journal}
  {\bibinfo  {journal} {The Astrophysical Journal}\ }\textbf {\bibinfo {volume}
  {900}},\ \bibinfo {pages} {L13} (\bibinfo {year}
  {2020}{\natexlab{b}})}\BibitemShut {NoStop}%
\bibitem [{\citenamefont {{Amaro-Seoane}}\ \emph {et~al.}(2017)\citenamefont
  {{Amaro-Seoane}}, \citenamefont {{Audley}}, \citenamefont {{Babak}},
  \citenamefont {{Baker}}, \citenamefont {{Barausse}}, \citenamefont
  {{Bender}}, \citenamefont {{Berti}}, \citenamefont {{Binetruy}},\ and\
  \citenamefont {et~al.}}]{Amaro-Seoane:17}%
  \BibitemOpen
  \bibfield  {author} {\bibinfo {author} {\bibfnamefont {P.}~\bibnamefont
  {{Amaro-Seoane}}}, \bibinfo {author} {\bibfnamefont {H.}~\bibnamefont
  {{Audley}}}, \bibinfo {author} {\bibfnamefont {S.}~\bibnamefont {{Babak}}},
  \bibinfo {author} {\bibfnamefont {J.}~\bibnamefont {{Baker}}}, \bibinfo
  {author} {\bibfnamefont {E.}~\bibnamefont {{Barausse}}}, \bibinfo {author}
  {\bibfnamefont {P.}~\bibnamefont {{Bender}}}, \bibinfo {author}
  {\bibfnamefont {E.}~\bibnamefont {{Berti}}}, \bibinfo {author} {\bibfnamefont
  {P.}~\bibnamefont {{Binetruy}}},\ and\ \bibinfo {author} {\bibnamefont
  {et~al.}},\ }\bibfield  {title} {\bibinfo {title} {{Laser Interferometer
  Space Antenna}},\ }\href@noop {} {\bibfield  {journal} {\bibinfo  {journal}
  {arXiv e-prints}\ ,\ \bibinfo {eid} {arXiv:1702.00786}} (\bibinfo {year}
  {2017})},\ \Eprint {https://arxiv.org/abs/1702.00786} {arXiv:1702.00786
  [astro-ph.IM]} \BibitemShut {NoStop}%
\bibitem [{\citenamefont {{Luo}}\ \emph {et~al.}(2016)\citenamefont {{Luo}},
  \citenamefont {{Chen}}, \citenamefont {{Duan}}, \citenamefont {{Gong}},
  \citenamefont {{Hu}}, \citenamefont {{Ji}}, \citenamefont {{Liu}},
  \citenamefont {{Mei}}, \citenamefont {{Milyukov}}, \citenamefont {{Sazhin}},
  \citenamefont {{Shao}}, \citenamefont {{Toth}}, \citenamefont {{Tu}},
  \citenamefont {{Wang}}, \citenamefont {{Wang}}, \citenamefont {{Yeh}},
  \citenamefont {{Zhan}}, \citenamefont {{Zhang}}, \citenamefont {{Zharov}},\
  and\ \citenamefont {{Zhou}}}]{Luo:16}%
  \BibitemOpen
  \bibfield  {author} {\bibinfo {author} {\bibfnamefont {J.}~\bibnamefont
  {{Luo}}}, \bibinfo {author} {\bibfnamefont {L.-S.}\ \bibnamefont {{Chen}}},
  \bibinfo {author} {\bibfnamefont {H.-Z.}\ \bibnamefont {{Duan}}}, \bibinfo
  {author} {\bibfnamefont {Y.-G.}\ \bibnamefont {{Gong}}}, \bibinfo {author}
  {\bibfnamefont {S.}~\bibnamefont {{Hu}}}, \bibinfo {author} {\bibfnamefont
  {J.}~\bibnamefont {{Ji}}}, \bibinfo {author} {\bibfnamefont {Q.}~\bibnamefont
  {{Liu}}}, \bibinfo {author} {\bibfnamefont {J.}~\bibnamefont {{Mei}}},
  \bibinfo {author} {\bibfnamefont {V.}~\bibnamefont {{Milyukov}}}, \bibinfo
  {author} {\bibfnamefont {M.}~\bibnamefont {{Sazhin}}}, \bibinfo {author}
  {\bibfnamefont {C.-G.}\ \bibnamefont {{Shao}}}, \bibinfo {author}
  {\bibfnamefont {V.~T.}\ \bibnamefont {{Toth}}}, \bibinfo {author}
  {\bibfnamefont {H.-B.}\ \bibnamefont {{Tu}}}, \bibinfo {author}
  {\bibfnamefont {Y.}~\bibnamefont {{Wang}}}, \bibinfo {author} {\bibfnamefont
  {Y.}~\bibnamefont {{Wang}}}, \bibinfo {author} {\bibfnamefont {H.-C.}\
  \bibnamefont {{Yeh}}}, \bibinfo {author} {\bibfnamefont {M.-S.}\ \bibnamefont
  {{Zhan}}}, \bibinfo {author} {\bibfnamefont {Y.}~\bibnamefont {{Zhang}}},
  \bibinfo {author} {\bibfnamefont {V.}~\bibnamefont {{Zharov}}},\ and\
  \bibinfo {author} {\bibfnamefont {Z.-B.}\ \bibnamefont {{Zhou}}},\ }\bibfield
   {title} {\bibinfo {title} {{TianQin: a space-borne gravitational wave
  detector}},\ }\href {https://doi.org/10.1088/0264-9381/33/3/035010}
  {\bibfield  {journal} {\bibinfo  {journal} {Classical and Quantum Gravity}\
  }\textbf {\bibinfo {volume} {33}},\ \bibinfo {eid} {035010} (\bibinfo {year}
  {2016})},\ \Eprint {https://arxiv.org/abs/1512.02076} {arXiv:1512.02076
  [astro-ph.IM]} \BibitemShut {NoStop}%
\bibitem [{\citenamefont {Hu}\ and\ \citenamefont {Wu}(2017)}]{Hu:17}%
  \BibitemOpen
  \bibfield  {author} {\bibinfo {author} {\bibfnamefont {W.-R.}\ \bibnamefont
  {Hu}}\ and\ \bibinfo {author} {\bibfnamefont {Y.-L.}\ \bibnamefont {Wu}},\
  }\bibfield  {title} {\bibinfo {title} {{The Taiji Program in Space for
  gravitational wave physics and the nature of gravity}},\ }\href
  {https://doi.org/10.1093/nsr/nwx116} {\bibfield  {journal} {\bibinfo
  {journal} {National Science Review}\ }\textbf {\bibinfo {volume} {4}},\
  \bibinfo {pages} {685} (\bibinfo {year} {2017})},\ \Eprint
  {https://arxiv.org/abs/https://academic.oup.com/nsr/article-pdf/4/5/685/31566708/nwx116.pdf}
  {https://academic.oup.com/nsr/article-pdf/4/5/685/31566708/nwx116.pdf}
  \BibitemShut {NoStop}%
\bibitem [{\citenamefont {{Nakamura}}\ \emph {et~al.}(2016)\citenamefont
  {{Nakamura}}, \citenamefont {{Ando}}, \citenamefont {{Kinugawa}},
  \citenamefont {{Nakano}}, \citenamefont {{Eda}}, \citenamefont {{Sato}},
  \citenamefont {{Musha}}, \citenamefont {{Akutsu}}, \citenamefont {{Tanaka}},
  \citenamefont {{Seto}}, \citenamefont {{Kanda}},\ and\ \citenamefont
  {{Itoh}}}]{Nakamura:16}%
  \BibitemOpen
  \bibfield  {author} {\bibinfo {author} {\bibfnamefont {T.}~\bibnamefont
  {{Nakamura}}}, \bibinfo {author} {\bibfnamefont {M.}~\bibnamefont {{Ando}}},
  \bibinfo {author} {\bibfnamefont {T.}~\bibnamefont {{Kinugawa}}}, \bibinfo
  {author} {\bibfnamefont {H.}~\bibnamefont {{Nakano}}}, \bibinfo {author}
  {\bibfnamefont {K.}~\bibnamefont {{Eda}}}, \bibinfo {author} {\bibfnamefont
  {S.}~\bibnamefont {{Sato}}}, \bibinfo {author} {\bibfnamefont
  {M.}~\bibnamefont {{Musha}}}, \bibinfo {author} {\bibfnamefont
  {T.}~\bibnamefont {{Akutsu}}}, \bibinfo {author} {\bibfnamefont
  {T.}~\bibnamefont {{Tanaka}}}, \bibinfo {author} {\bibfnamefont
  {N.}~\bibnamefont {{Seto}}}, \bibinfo {author} {\bibfnamefont
  {N.}~\bibnamefont {{Kanda}}},\ and\ \bibinfo {author} {\bibfnamefont
  {Y.}~\bibnamefont {{Itoh}}},\ }\bibfield  {title} {\bibinfo {title}
  {{Pre-DECIGO can get the smoking gun to decide the astrophysical or
  cosmological origin of GW150914-like binary black holes}},\ }\href
  {https://doi.org/10.1093/ptep/ptw127} {\bibfield  {journal} {\bibinfo
  {journal} {Progress of Theoretical and Experimental Physics}\ }\textbf
  {\bibinfo {volume} {2016}},\ \bibinfo {eid} {093E01} (\bibinfo {year}
  {2016})},\ \Eprint {https://arxiv.org/abs/1607.00897} {arXiv:1607.00897
  [astro-ph.HE]} \BibitemShut {NoStop}%
\bibitem [{\citenamefont {{Kawamura}}\ \emph {et~al.}(2020)\citenamefont
  {{Kawamura}}, \citenamefont {{Ando}}, \citenamefont {{Seto}}, \citenamefont
  {{Sato}}, \citenamefont {{Musha}}, \citenamefont {{Kawano}}, \citenamefont
  {{Yokoyama}}, \citenamefont {{Tanaka}},\ and\ \citenamefont
  {et~al.}}]{Kawamura:20}%
  \BibitemOpen
  \bibfield  {author} {\bibinfo {author} {\bibfnamefont {S.}~\bibnamefont
  {{Kawamura}}}, \bibinfo {author} {\bibfnamefont {M.}~\bibnamefont {{Ando}}},
  \bibinfo {author} {\bibfnamefont {N.}~\bibnamefont {{Seto}}}, \bibinfo
  {author} {\bibfnamefont {S.}~\bibnamefont {{Sato}}}, \bibinfo {author}
  {\bibfnamefont {M.}~\bibnamefont {{Musha}}}, \bibinfo {author} {\bibfnamefont
  {I.}~\bibnamefont {{Kawano}}}, \bibinfo {author} {\bibfnamefont
  {J.}~\bibnamefont {{Yokoyama}}}, \bibinfo {author} {\bibfnamefont
  {T.}~\bibnamefont {{Tanaka}}},\ and\ \bibinfo {author} {\bibnamefont
  {et~al.}},\ }\bibfield  {title} {\bibinfo {title} {{Current status of space
  gravitational wave antenna DECIGO and B-DECIGO}},\ }\href@noop {} {\bibfield
  {journal} {\bibinfo  {journal} {arXiv e-prints}\ ,\ \bibinfo {eid}
  {arXiv:2006.13545}} (\bibinfo {year} {2020})},\ \Eprint
  {https://arxiv.org/abs/2006.13545} {arXiv:2006.13545 [gr-qc]} \BibitemShut
  {NoStop}%
\bibitem [{\citenamefont {{Arca Sedda}}\ \emph {et~al.}(2019)\citenamefont
  {{Arca Sedda}}, \citenamefont {{Berry}}, \citenamefont {{Jani}},
  \citenamefont {{Amaro-Seoane}}, \citenamefont {{Auclair}}, \citenamefont
  {{Baird}}, \citenamefont {{Baker}}, \citenamefont {{Berti}},\ and\
  \citenamefont {et~al.}}]{Sedda:19}%
  \BibitemOpen
  \bibfield  {author} {\bibinfo {author} {\bibfnamefont {M.}~\bibnamefont
  {{Arca Sedda}}}, \bibinfo {author} {\bibfnamefont {C.~P.~L.}\ \bibnamefont
  {{Berry}}}, \bibinfo {author} {\bibfnamefont {K.}~\bibnamefont {{Jani}}},
  \bibinfo {author} {\bibfnamefont {P.}~\bibnamefont {{Amaro-Seoane}}},
  \bibinfo {author} {\bibfnamefont {P.}~\bibnamefont {{Auclair}}}, \bibinfo
  {author} {\bibfnamefont {J.}~\bibnamefont {{Baird}}}, \bibinfo {author}
  {\bibfnamefont {T.}~\bibnamefont {{Baker}}}, \bibinfo {author} {\bibfnamefont
  {E.}~\bibnamefont {{Berti}}},\ and\ \bibinfo {author} {\bibnamefont
  {et~al.}},\ }\bibfield  {title} {\bibinfo {title} {{The Missing Link in
  Gravitational-Wave Astronomy: Discoveries waiting in the decihertz range}},\
  }\href@noop {} {\bibfield  {journal} {\bibinfo  {journal} {arXiv e-prints}\
  ,\ \bibinfo {eid} {arXiv:1908.11375}} (\bibinfo {year} {2019})},\ \Eprint
  {https://arxiv.org/abs/1908.11375} {arXiv:1908.11375 [gr-qc]} \BibitemShut
  {NoStop}%
\bibitem [{\citenamefont {Kuns}\ \emph {et~al.}(2020)\citenamefont {Kuns},
  \citenamefont {Yu}, \citenamefont {Chen},\ and\ \citenamefont
  {Adhikari}}]{Kuns:19}%
  \BibitemOpen
  \bibfield  {author} {\bibinfo {author} {\bibfnamefont {K.~A.}\ \bibnamefont
  {Kuns}}, \bibinfo {author} {\bibfnamefont {H.}~\bibnamefont {Yu}}, \bibinfo
  {author} {\bibfnamefont {Y.}~\bibnamefont {Chen}},\ and\ \bibinfo {author}
  {\bibfnamefont {R.~X.}\ \bibnamefont {Adhikari}},\ }\bibfield  {title}
  {\bibinfo {title} {Astrophysics and cosmology with a decihertz
  gravitational-wave detector: Tiango},\ }\href
  {https://doi.org/10.1103/PhysRevD.102.043001} {\bibfield  {journal} {\bibinfo
   {journal} {Phys. Rev. D}\ }\textbf {\bibinfo {volume} {102}},\ \bibinfo
  {pages} {043001} (\bibinfo {year} {2020})}\BibitemShut {NoStop}%
\bibitem [{\citenamefont {{Damour}}\ and\ \citenamefont
  {{Taylor}}(1992)}]{Damour:92}%
  \BibitemOpen
  \bibfield  {author} {\bibinfo {author} {\bibfnamefont {T.}~\bibnamefont
  {{Damour}}}\ and\ \bibinfo {author} {\bibfnamefont {J.~H.}\ \bibnamefont
  {{Taylor}}},\ }\bibfield  {title} {\bibinfo {title} {{Strong-field tests of
  relativistic gravity and binary pulsars}},\ }\href
  {https://doi.org/10.1103/PhysRevD.45.1840} {\bibfield  {journal} {\bibinfo
  {journal} {\prd}\ }\textbf {\bibinfo {volume} {45}},\ \bibinfo {pages} {1840}
  (\bibinfo {year} {1992})}\BibitemShut {NoStop}%
\bibitem [{\citenamefont {{Amaro-Seoane}}(2018)}]{Amaro-Seoane:18}%
  \BibitemOpen
  \bibfield  {author} {\bibinfo {author} {\bibfnamefont {P.}~\bibnamefont
  {{Amaro-Seoane}}},\ }\bibfield  {title} {\bibinfo {title} {{Relativistic
  dynamics and extreme mass ratio inspirals}},\ }\href
  {https://doi.org/10.1007/s41114-018-0013-8} {\bibfield  {journal} {\bibinfo
  {journal} {Living Reviews in Relativity}\ }\textbf {\bibinfo {volume} {21}},\
  \bibinfo {eid} {4} (\bibinfo {year} {2018})},\ \Eprint
  {https://arxiv.org/abs/1205.5240} {arXiv:1205.5240 [astro-ph.CO]}
  \BibitemShut {NoStop}%
\bibitem [{\citenamefont {{Amaro-Seoane}}(2019)}]{Amaro-Seoane:19}%
  \BibitemOpen
  \bibfield  {author} {\bibinfo {author} {\bibfnamefont {P.}~\bibnamefont
  {{Amaro-Seoane}}},\ }\bibfield  {title} {\bibinfo {title} {{Extremely large
  mass-ratio inspirals}},\ }\href {https://doi.org/10.1103/PhysRevD.99.123025}
  {\bibfield  {journal} {\bibinfo  {journal} {\prd}\ }\textbf {\bibinfo
  {volume} {99}},\ \bibinfo {eid} {123025} (\bibinfo {year} {2019})},\ \Eprint
  {https://arxiv.org/abs/1903.10871} {arXiv:1903.10871 [astro-ph.GA]}
  \BibitemShut {NoStop}%
\bibitem [{\citenamefont {{Han}}\ and\ \citenamefont {{Chen}}(2019)}]{Han:19}%
  \BibitemOpen
  \bibfield  {author} {\bibinfo {author} {\bibfnamefont {W.-B.}\ \bibnamefont
  {{Han}}}\ and\ \bibinfo {author} {\bibfnamefont {X.}~\bibnamefont {{Chen}}},\
  }\bibfield  {title} {\bibinfo {title} {{Testing general relativity using
  binary extreme-mass-ratio inspirals}},\ }\href
  {https://doi.org/10.1093/mnrasl/slz021} {\bibfield  {journal} {\bibinfo
  {journal} {\mnras}\ }\textbf {\bibinfo {volume} {485}},\ \bibinfo {pages}
  {L29} (\bibinfo {year} {2019})},\ \Eprint {https://arxiv.org/abs/1801.07060}
  {arXiv:1801.07060 [gr-qc]} \BibitemShut {NoStop}%
\bibitem [{\citenamefont {{Han}}\ \emph {et~al.}(2020)\citenamefont {{Han}},
  \citenamefont {{Zhong}}, \citenamefont {{Chen}},\ and\ \citenamefont
  {{Xin}}}]{Han:20}%
  \BibitemOpen
  \bibfield  {author} {\bibinfo {author} {\bibfnamefont {W.-B.}\ \bibnamefont
  {{Han}}}, \bibinfo {author} {\bibfnamefont {X.-Y.}\ \bibnamefont {{Zhong}}},
  \bibinfo {author} {\bibfnamefont {X.}~\bibnamefont {{Chen}}},\ and\ \bibinfo
  {author} {\bibfnamefont {S.}~\bibnamefont {{Xin}}},\ }\bibfield  {title}
  {\bibinfo {title} {{Very extreme mass-ratio bursts in the Galaxy and
  neighbouring galaxies in relation to space-borne detectors}},\ }\href
  {https://doi.org/10.1093/mnrasl/slaa115} {\bibfield  {journal} {\bibinfo
  {journal} {\mnras}\ }\textbf {\bibinfo {volume} {498}},\ \bibinfo {pages}
  {L61} (\bibinfo {year} {2020})}\BibitemShut {NoStop}%
\bibitem [{Note1()}]{Note1}%
  \BibitemOpen
  \bibinfo {note} {For future convenience, we will refer to the orbit of the
  stellar-mass BH binary as the inner orbit, and its center of mass orbiting
  the SMBH as the outer orbit.}\BibitemShut {Stop}%
\bibitem [{\citenamefont {{Inayoshi}}\ \emph {et~al.}(2017)\citenamefont
  {{Inayoshi}}, \citenamefont {{Tamanini}}, \citenamefont {{Caprini}},\ and\
  \citenamefont {{Haiman}}}]{Inayoshi:17}%
  \BibitemOpen
  \bibfield  {author} {\bibinfo {author} {\bibfnamefont {K.}~\bibnamefont
  {{Inayoshi}}}, \bibinfo {author} {\bibfnamefont {N.}~\bibnamefont
  {{Tamanini}}}, \bibinfo {author} {\bibfnamefont {C.}~\bibnamefont
  {{Caprini}}},\ and\ \bibinfo {author} {\bibfnamefont {Z.}~\bibnamefont
  {{Haiman}}},\ }\bibfield  {title} {\bibinfo {title} {{Probing stellar binary
  black hole formation in galactic nuclei via the imprint of their center of
  mass acceleration on their gravitational wave signal}},\ }\href
  {https://doi.org/10.1103/PhysRevD.96.063014} {\bibfield  {journal} {\bibinfo
  {journal} {\prd}\ }\textbf {\bibinfo {volume} {96}},\ \bibinfo {eid} {063014}
  (\bibinfo {year} {2017})},\ \Eprint {https://arxiv.org/abs/1702.06529}
  {arXiv:1702.06529 [astro-ph.HE]} \BibitemShut {NoStop}%
\bibitem [{\citenamefont {{Randall}}\ and\ \citenamefont
  {{Xianyu}}(2019)}]{Randall:19}%
  \BibitemOpen
  \bibfield  {author} {\bibinfo {author} {\bibfnamefont {L.}~\bibnamefont
  {{Randall}}}\ and\ \bibinfo {author} {\bibfnamefont {Z.-Z.}\ \bibnamefont
  {{Xianyu}}},\ }\bibfield  {title} {\bibinfo {title} {{A Direct Probe of Mass
  Density near Inspiraling Binary Black Holes}},\ }\href
  {https://doi.org/10.3847/1538-4357/ab20c6} {\bibfield  {journal} {\bibinfo
  {journal} {\apj}\ }\textbf {\bibinfo {volume} {878}},\ \bibinfo {eid} {75}
  (\bibinfo {year} {2019})},\ \Eprint {https://arxiv.org/abs/1805.05335}
  {arXiv:1805.05335 [gr-qc]} \BibitemShut {NoStop}%
\bibitem [{\citenamefont {{Will}}(2018)}]{Will:18}%
  \BibitemOpen
  \bibfield  {author} {\bibinfo {author} {\bibfnamefont {C.~M.}\ \bibnamefont
  {{Will}}},\ }\bibfield  {title} {\bibinfo {title} {{New General Relativistic
  Contribution to Mercury's Perihelion Advance}},\ }\href
  {https://doi.org/10.1103/PhysRevLett.120.191101} {\bibfield  {journal}
  {\bibinfo  {journal} {\prl}\ }\textbf {\bibinfo {volume} {120}},\ \bibinfo
  {eid} {191101} (\bibinfo {year} {2018})},\ \Eprint
  {https://arxiv.org/abs/1802.05304} {arXiv:1802.05304 [gr-qc]} \BibitemShut
  {NoStop}%
\bibitem [{\citenamefont {{Liu}}\ \emph {et~al.}(2019)\citenamefont {{Liu}},
  \citenamefont {{Lai}},\ and\ \citenamefont {{Wang}}}]{Liu:19}%
  \BibitemOpen
  \bibfield  {author} {\bibinfo {author} {\bibfnamefont {B.}~\bibnamefont
  {{Liu}}}, \bibinfo {author} {\bibfnamefont {D.}~\bibnamefont {{Lai}}},\ and\
  \bibinfo {author} {\bibfnamefont {Y.-H.}\ \bibnamefont {{Wang}}},\ }\bibfield
   {title} {\bibinfo {title} {{Binary Mergers near a Supermassive Black Hole:
  Relativistic Effects in Triples}},\ }\href
  {https://doi.org/10.3847/2041-8213/ab40c0} {\bibfield  {journal} {\bibinfo
  {journal} {\apjl}\ }\textbf {\bibinfo {volume} {883}},\ \bibinfo {eid} {L7}
  (\bibinfo {year} {2019})},\ \Eprint {https://arxiv.org/abs/1906.07726}
  {arXiv:1906.07726 [astro-ph.HE]} \BibitemShut {NoStop}%
\bibitem [{\citenamefont {{Yu}}\ \emph {et~al.}(2020)\citenamefont {{Yu}},
  \citenamefont {{Ma}}, \citenamefont {{Giesler}},\ and\ \citenamefont
  {{Chen}}}]{Yu:20b}%
  \BibitemOpen
  \bibfield  {author} {\bibinfo {author} {\bibfnamefont {H.}~\bibnamefont
  {{Yu}}}, \bibinfo {author} {\bibfnamefont {S.}~\bibnamefont {{Ma}}}, \bibinfo
  {author} {\bibfnamefont {M.}~\bibnamefont {{Giesler}}},\ and\ \bibinfo
  {author} {\bibfnamefont {Y.}~\bibnamefont {{Chen}}},\ }\bibfield  {title}
  {\bibinfo {title} {{Spin and eccentricity evolution in triple systems: From
  the Lidov-Kozai interaction to the final merger of the inner binary}},\
  }\href {https://doi.org/10.1103/PhysRevD.102.123009} {\bibfield  {journal}
  {\bibinfo  {journal} {\prd}\ }\textbf {\bibinfo {volume} {102}},\ \bibinfo
  {eid} {123009} (\bibinfo {year} {2020})},\ \Eprint
  {https://arxiv.org/abs/2007.12978} {arXiv:2007.12978 [gr-qc]} \BibitemShut
  {NoStop}%
\bibitem [{\citenamefont {{Kiseleva}}\ \emph {et~al.}(1996)\citenamefont
  {{Kiseleva}}, \citenamefont {{Aarseth}}, \citenamefont {{Eggleton}},\ and\
  \citenamefont {{de La Fuente Marcos}}}]{Kiseleva:96}%
  \BibitemOpen
  \bibfield  {author} {\bibinfo {author} {\bibfnamefont {L.~G.}\ \bibnamefont
  {{Kiseleva}}}, \bibinfo {author} {\bibfnamefont {S.~J.}\ \bibnamefont
  {{Aarseth}}}, \bibinfo {author} {\bibfnamefont {P.~P.}\ \bibnamefont
  {{Eggleton}}},\ and\ \bibinfo {author} {\bibfnamefont {R.}~\bibnamefont {{de
  La Fuente Marcos}}},\ }\bibfield  {title} {\bibinfo {title} {{Formation and
  Evolution of Hierarchical Triple Systems in Open Clusters}},\ }in\ \href@noop
  {} {\emph {\bibinfo {booktitle} {The Origins, Evolution, and Destinies of
  Binary Stars in Clusters}}},\ \bibinfo {series} {Astronomical Society of the
  Pacific Conference Series}, Vol.~\bibinfo {volume} {90},\ \bibinfo {editor}
  {edited by\ \bibinfo {editor} {\bibfnamefont {E.~F.}\ \bibnamefont
  {{Milone}}}\ and\ \bibinfo {editor} {\bibfnamefont {J.~C.}\ \bibnamefont
  {{Mermilliod}}}}\ (\bibinfo {year} {1996})\ p.\ \bibinfo {pages}
  {433}\BibitemShut {NoStop}%
\bibitem [{\citenamefont {{Bellovary}}\ \emph {et~al.}(2016)\citenamefont
  {{Bellovary}}, \citenamefont {{Mac Low}}, \citenamefont {{McKernan}},\ and\
  \citenamefont {{Ford}}}]{Bellovary:16}%
  \BibitemOpen
  \bibfield  {author} {\bibinfo {author} {\bibfnamefont {J.~M.}\ \bibnamefont
  {{Bellovary}}}, \bibinfo {author} {\bibfnamefont {M.-M.}\ \bibnamefont {{Mac
  Low}}}, \bibinfo {author} {\bibfnamefont {B.}~\bibnamefont {{McKernan}}},\
  and\ \bibinfo {author} {\bibfnamefont {K.~E.~S.}\ \bibnamefont {{Ford}}},\
  }\bibfield  {title} {\bibinfo {title} {{Migration Traps in Disks around
  Supermassive Black Holes}},\ }\href
  {https://doi.org/10.3847/2041-8205/819/2/L17} {\bibfield  {journal} {\bibinfo
   {journal} {\apjl}\ }\textbf {\bibinfo {volume} {819}},\ \bibinfo {eid} {L17}
  (\bibinfo {year} {2016})},\ \Eprint {https://arxiv.org/abs/1511.00005}
  {arXiv:1511.00005 [astro-ph.GA]} \BibitemShut {NoStop}%
\bibitem [{\citenamefont {{VanLandingham}}\ \emph {et~al.}(2016)\citenamefont
  {{VanLandingham}}, \citenamefont {{Miller}}, \citenamefont {{Hamilton}},\
  and\ \citenamefont {{Richardson}}}]{VanLandingham:16}%
  \BibitemOpen
  \bibfield  {author} {\bibinfo {author} {\bibfnamefont {J.~H.}\ \bibnamefont
  {{VanLandingham}}}, \bibinfo {author} {\bibfnamefont {M.~C.}\ \bibnamefont
  {{Miller}}}, \bibinfo {author} {\bibfnamefont {D.~P.}\ \bibnamefont
  {{Hamilton}}},\ and\ \bibinfo {author} {\bibfnamefont {D.~C.}\ \bibnamefont
  {{Richardson}}},\ }\bibfield  {title} {\bibinfo {title} {{The Role of the
  Kozai--Lidov Mechanism in Black Hole Binary Mergers in Galactic Centers}},\
  }\href {https://doi.org/10.3847/0004-637X/828/2/77} {\bibfield  {journal}
  {\bibinfo  {journal} {\apj}\ }\textbf {\bibinfo {volume} {828}},\ \bibinfo
  {eid} {77} (\bibinfo {year} {2016})},\ \Eprint
  {https://arxiv.org/abs/1604.04948} {arXiv:1604.04948 [astro-ph.HE]}
  \BibitemShut {NoStop}%
\bibitem [{\citenamefont {{Flanagan}}\ and\ \citenamefont
  {{Hughes}}(1998)}]{Flanagan:98}%
  \BibitemOpen
  \bibfield  {author} {\bibinfo {author} {\bibfnamefont {{\'E}.~{\'E}.}\
  \bibnamefont {{Flanagan}}}\ and\ \bibinfo {author} {\bibfnamefont {S.~A.}\
  \bibnamefont {{Hughes}}},\ }\bibfield  {title} {\bibinfo {title} {{Measuring
  gravitational waves from binary black hole coalescences. I. Signal to noise
  for inspiral, merger, and ringdown}},\ }\href
  {https://doi.org/10.1103/PhysRevD.57.4535} {\bibfield  {journal} {\bibinfo
  {journal} {\prd}\ }\textbf {\bibinfo {volume} {57}},\ \bibinfo {pages} {4535}
  (\bibinfo {year} {1998})},\ \Eprint {https://arxiv.org/abs/gr-qc/9701039}
  {arXiv:gr-qc/9701039 [gr-qc]} \BibitemShut {NoStop}%
\bibitem [{\citenamefont {{Peterson}}(2014)}]{Peterson:14}%
  \BibitemOpen
  \bibfield  {author} {\bibinfo {author} {\bibfnamefont {B.~M.}\ \bibnamefont
  {{Peterson}}},\ }\bibfield  {title} {\bibinfo {title} {{Measuring the Masses
  of Supermassive Black Holes}},\ }\href
  {https://doi.org/10.1007/s11214-013-9987-4} {\bibfield  {journal} {\bibinfo
  {journal} {\ssr}\ }\textbf {\bibinfo {volume} {183}},\ \bibinfo {pages} {253}
  (\bibinfo {year} {2014})}\BibitemShut {NoStop}%
\bibitem [{\citenamefont {{Hughes}}(2002)}]{Hughes:02}%
  \BibitemOpen
  \bibfield  {author} {\bibinfo {author} {\bibfnamefont {S.~A.}\ \bibnamefont
  {{Hughes}}},\ }\bibfield  {title} {\bibinfo {title} {{Untangling the merger
  history of massive black holes with LISA}},\ }\href
  {https://doi.org/10.1046/j.1365-8711.2002.05247.x} {\bibfield  {journal}
  {\bibinfo  {journal} {\mnras}\ }\textbf {\bibinfo {volume} {331}},\ \bibinfo
  {pages} {805} (\bibinfo {year} {2002})},\ \Eprint
  {https://arxiv.org/abs/astro-ph/0108483} {arXiv:astro-ph/0108483 [astro-ph]}
  \BibitemShut {NoStop}%
\bibitem [{\citenamefont {{Gair}}\ \emph {et~al.}(2004)\citenamefont {{Gair}},
  \citenamefont {{Barack}}, \citenamefont {{Creighton}}, \citenamefont
  {{Cutler}}, \citenamefont {{Larson}}, \citenamefont {{Phinney}},\ and\
  \citenamefont {{Vallisneri}}}]{Gair:04}%
  \BibitemOpen
  \bibfield  {author} {\bibinfo {author} {\bibfnamefont {J.~R.}\ \bibnamefont
  {{Gair}}}, \bibinfo {author} {\bibfnamefont {L.}~\bibnamefont {{Barack}}},
  \bibinfo {author} {\bibfnamefont {T.}~\bibnamefont {{Creighton}}}, \bibinfo
  {author} {\bibfnamefont {C.}~\bibnamefont {{Cutler}}}, \bibinfo {author}
  {\bibfnamefont {S.~L.}\ \bibnamefont {{Larson}}}, \bibinfo {author}
  {\bibfnamefont {E.~S.}\ \bibnamefont {{Phinney}}},\ and\ \bibinfo {author}
  {\bibfnamefont {M.}~\bibnamefont {{Vallisneri}}},\ }\bibfield  {title}
  {\bibinfo {title} {{Event rate estimates for LISA extreme mass ratio capture
  sources}},\ }\href {https://doi.org/10.1088/0264-9381/21/20/003} {\bibfield
  {journal} {\bibinfo  {journal} {Classical and Quantum Gravity}\ }\textbf
  {\bibinfo {volume} {21}},\ \bibinfo {pages} {S1595} (\bibinfo {year}
  {2004})},\ \Eprint {https://arxiv.org/abs/gr-qc/0405137} {arXiv:gr-qc/0405137
  [gr-qc]} \BibitemShut {NoStop}%
\bibitem [{\citenamefont {{Gair}}\ \emph {et~al.}(2017)\citenamefont {{Gair}},
  \citenamefont {{Babak}}, \citenamefont {{Sesana}}, \citenamefont
  {{Amaro-Seoane}}, \citenamefont {{Barausse}}, \citenamefont {{Berry}},
  \citenamefont {{Berti}},\ and\ \citenamefont {{Sopuerta}}}]{Gair:17}%
  \BibitemOpen
  \bibfield  {author} {\bibinfo {author} {\bibfnamefont {J.~R.}\ \bibnamefont
  {{Gair}}}, \bibinfo {author} {\bibfnamefont {S.}~\bibnamefont {{Babak}}},
  \bibinfo {author} {\bibfnamefont {A.}~\bibnamefont {{Sesana}}}, \bibinfo
  {author} {\bibfnamefont {P.}~\bibnamefont {{Amaro-Seoane}}}, \bibinfo
  {author} {\bibfnamefont {E.}~\bibnamefont {{Barausse}}}, \bibinfo {author}
  {\bibfnamefont {C.~P.~L.}\ \bibnamefont {{Berry}}}, \bibinfo {author}
  {\bibfnamefont {E.}~\bibnamefont {{Berti}}},\ and\ \bibinfo {author}
  {\bibfnamefont {C.}~\bibnamefont {{Sopuerta}}},\ }\bibfield  {title}
  {\bibinfo {title} {{Prospects for observing extreme-mass-ratio inspirals with
  LISA}},\ }in\ \href {https://doi.org/10.1088/1742-6596/840/1/012021} {\emph
  {\bibinfo {booktitle} {Journal of Physics Conference Series}}},\ \bibinfo
  {series} {Journal of Physics Conference Series}, Vol.\ \bibinfo {volume}
  {840}\ (\bibinfo {year} {2017})\ p.\ \bibinfo {pages} {012021},\ \Eprint
  {https://arxiv.org/abs/1704.00009} {arXiv:1704.00009 [astro-ph.GA]}
  \BibitemShut {NoStop}%
\bibitem [{\citenamefont {{Fang}}\ \emph {et~al.}(2019)\citenamefont {{Fang}},
  \citenamefont {{Chen}},\ and\ \citenamefont {{Huang}}}]{Fang:19}%
  \BibitemOpen
  \bibfield  {author} {\bibinfo {author} {\bibfnamefont {Y.}~\bibnamefont
  {{Fang}}}, \bibinfo {author} {\bibfnamefont {X.}~\bibnamefont {{Chen}}},\
  and\ \bibinfo {author} {\bibfnamefont {Q.-G.}\ \bibnamefont {{Huang}}},\
  }\bibfield  {title} {\bibinfo {title} {{Impact of a Spinning Supermassive
  Black Hole on the Orbit and Gravitational Waves of a Nearby Compact
  Binary}},\ }\href {https://doi.org/10.3847/1538-4357/ab510e} {\bibfield
  {journal} {\bibinfo  {journal} {\apj}\ }\textbf {\bibinfo {volume} {887}},\
  \bibinfo {eid} {210} (\bibinfo {year} {2019})},\ \Eprint
  {https://arxiv.org/abs/1908.01443} {arXiv:1908.01443 [astro-ph.HE]}
  \BibitemShut {NoStop}%
\bibitem [{\citenamefont {{Ostriker}}(1999)}]{Ostriker:99}%
  \BibitemOpen
  \bibfield  {author} {\bibinfo {author} {\bibfnamefont {E.~C.}\ \bibnamefont
  {{Ostriker}}},\ }\bibfield  {title} {\bibinfo {title} {{Dynamical Friction in
  a Gaseous Medium}},\ }\href {https://doi.org/10.1086/306858} {\bibfield
  {journal} {\bibinfo  {journal} {\apj}\ }\textbf {\bibinfo {volume} {513}},\
  \bibinfo {pages} {252} (\bibinfo {year} {1999})},\ \Eprint
  {https://arxiv.org/abs/astro-ph/9810324} {arXiv:astro-ph/9810324 [astro-ph]}
  \BibitemShut {NoStop}%
\bibitem [{\citenamefont {{Baruteau}}\ \emph {et~al.}(2011)\citenamefont
  {{Baruteau}}, \citenamefont {{Cuadra}},\ and\ \citenamefont
  {{Lin}}}]{Baruteau:11}%
  \BibitemOpen
  \bibfield  {author} {\bibinfo {author} {\bibfnamefont {C.}~\bibnamefont
  {{Baruteau}}}, \bibinfo {author} {\bibfnamefont {J.}~\bibnamefont
  {{Cuadra}}},\ and\ \bibinfo {author} {\bibfnamefont {D.~N.~C.}\ \bibnamefont
  {{Lin}}},\ }\bibfield  {title} {\bibinfo {title} {{Binaries Migrating in a
  Gaseous Disk: Where are the Galactic Center Binaries?}},\ }\href
  {https://doi.org/10.1088/0004-637X/726/1/28} {\bibfield  {journal} {\bibinfo
  {journal} {\apj}\ }\textbf {\bibinfo {volume} {726}},\ \bibinfo {eid} {28}
  (\bibinfo {year} {2011})},\ \Eprint {https://arxiv.org/abs/1011.0360}
  {arXiv:1011.0360 [astro-ph.GA]} \BibitemShut {NoStop}%
\bibitem [{\citenamefont {{Antoni}}\ \emph {et~al.}(2019)\citenamefont
  {{Antoni}}, \citenamefont {{MacLeod}},\ and\ \citenamefont
  {{Ramirez-Ruiz}}}]{Antoni:19}%
  \BibitemOpen
  \bibfield  {author} {\bibinfo {author} {\bibfnamefont {A.}~\bibnamefont
  {{Antoni}}}, \bibinfo {author} {\bibfnamefont {M.}~\bibnamefont
  {{MacLeod}}},\ and\ \bibinfo {author} {\bibfnamefont {E.}~\bibnamefont
  {{Ramirez-Ruiz}}},\ }\bibfield  {title} {\bibinfo {title} {{The Evolution of
  Binaries in a Gaseous Medium: Three-dimensional Simulations of Binary
  Bondi-Hoyle-Lyttleton Accretion}},\ }\href
  {https://doi.org/10.3847/1538-4357/ab3466} {\bibfield  {journal} {\bibinfo
  {journal} {\apj}\ }\textbf {\bibinfo {volume} {884}},\ \bibinfo {eid} {22}
  (\bibinfo {year} {2019})},\ \Eprint {https://arxiv.org/abs/1901.07572}
  {arXiv:1901.07572 [astro-ph.HE]} \BibitemShut {NoStop}%
\bibitem [{\citenamefont {{Chen}}\ \emph {et~al.}(2020)\citenamefont {{Chen}},
  \citenamefont {{Xuan}},\ and\ \citenamefont {{Peng}}}]{Chen:20}%
  \BibitemOpen
  \bibfield  {author} {\bibinfo {author} {\bibfnamefont {X.}~\bibnamefont
  {{Chen}}}, \bibinfo {author} {\bibfnamefont {Z.-Y.}\ \bibnamefont {{Xuan}}},\
  and\ \bibinfo {author} {\bibfnamefont {P.}~\bibnamefont {{Peng}}},\
  }\bibfield  {title} {\bibinfo {title} {{Fake Massive Black Holes in the
  Milli-Hertz Gravitational-wave Band}},\ }\href
  {https://doi.org/10.3847/1538-4357/ab919f} {\bibfield  {journal} {\bibinfo
  {journal} {\apj}\ }\textbf {\bibinfo {volume} {896}},\ \bibinfo {eid} {171}
  (\bibinfo {year} {2020})},\ \Eprint {https://arxiv.org/abs/2003.08639}
  {arXiv:2003.08639 [astro-ph.HE]} \BibitemShut {NoStop}%
\bibitem [{\citenamefont {{Derdzinski}}\ \emph {et~al.}(2020)\citenamefont
  {{Derdzinski}}, \citenamefont {{D'Orazio}}, \citenamefont {{Duffell}},
  \citenamefont {{Haiman}},\ and\ \citenamefont {{Macfadyen}}}]{Derdzinski:20}%
  \BibitemOpen
  \bibfield  {author} {\bibinfo {author} {\bibfnamefont {A.}~\bibnamefont
  {{Derdzinski}}}, \bibinfo {author} {\bibfnamefont {D.}~\bibnamefont
  {{D'Orazio}}}, \bibinfo {author} {\bibfnamefont {P.}~\bibnamefont
  {{Duffell}}}, \bibinfo {author} {\bibfnamefont {Z.}~\bibnamefont
  {{Haiman}}},\ and\ \bibinfo {author} {\bibfnamefont {A.}~\bibnamefont
  {{Macfadyen}}},\ }\bibfield  {title} {\bibinfo {title} {{Evolution of gas
  disc-embedded intermediate mass ratio inspirals in the LISA band}},\
  }\href@noop {} {\bibfield  {journal} {\bibinfo  {journal} {arXiv e-prints}\
  ,\ \bibinfo {eid} {arXiv:2005.11333}} (\bibinfo {year} {2020})},\ \Eprint
  {https://arxiv.org/abs/2005.11333} {arXiv:2005.11333 [astro-ph.HE]}
  \BibitemShut {NoStop}%
\bibitem [{\citenamefont {{G{\"u}ltekin}}\ \emph {et~al.}(2004)\citenamefont
  {{G{\"u}ltekin}}, \citenamefont {{Miller}},\ and\ \citenamefont
  {{Hamilton}}}]{Gultekin:04}%
  \BibitemOpen
  \bibfield  {author} {\bibinfo {author} {\bibfnamefont {K.}~\bibnamefont
  {{G{\"u}ltekin}}}, \bibinfo {author} {\bibfnamefont {M.~C.}\ \bibnamefont
  {{Miller}}},\ and\ \bibinfo {author} {\bibfnamefont {D.~P.}\ \bibnamefont
  {{Hamilton}}},\ }\bibfield  {title} {\bibinfo {title} {{Growth of
  Intermediate-Mass Black Holes in Globular Clusters}},\ }\href
  {https://doi.org/10.1086/424809} {\bibfield  {journal} {\bibinfo  {journal}
  {\apj}\ }\textbf {\bibinfo {volume} {616}},\ \bibinfo {pages} {221} (\bibinfo
  {year} {2004})},\ \Eprint {https://arxiv.org/abs/astro-ph/0402532}
  {arXiv:astro-ph/0402532 [astro-ph]} \BibitemShut {NoStop}%
\bibitem [{Note2()}]{Note2}%
  \BibitemOpen
  \bibinfo {note} {One can show that from the source frame to the detector
  frame, $M_3\to (1+z)M_3$ and $a_\protect \text {o}\to (1+z)a_\protect \text
  {o}$ due to a constant redshift $z$. Note $z$ may include both the
  cosmological redshift ($\sim 0.2$ at $1\protect \tmspace +\thinmuskip
  {.1667em}{\protect \rm GPc}$) and that due to the gravitational potential of
  $M_3$ ($\lesssim 0.01$ for typical sources at $a_\protect \text {o}\gtrsim
  100 M_3$).}\BibitemShut {Stop}%
\bibitem [{\citenamefont {{Cutler}}(1998)}]{Cutler:98}%
  \BibitemOpen
  \bibfield  {author} {\bibinfo {author} {\bibfnamefont {C.}~\bibnamefont
  {{Cutler}}},\ }\bibfield  {title} {\bibinfo {title} {{Angular resolution of
  the LISA gravitational wave detector}},\ }\href
  {https://doi.org/10.1103/PhysRevD.57.7089} {\bibfield  {journal} {\bibinfo
  {journal} {\prd}\ }\textbf {\bibinfo {volume} {57}},\ \bibinfo {pages} {7089}
  (\bibinfo {year} {1998})},\ \Eprint {https://arxiv.org/abs/gr-qc/9703068}
  {arXiv:gr-qc/9703068 [gr-qc]} \BibitemShut {NoStop}%
\bibitem [{\citenamefont {{Apostolatos}}\ \emph {et~al.}(1994)\citenamefont
  {{Apostolatos}}, \citenamefont {{Cutler}}, \citenamefont {{Sussman}},\ and\
  \citenamefont {{Thorne}}}]{Apostolatos:94}%
  \BibitemOpen
  \bibfield  {author} {\bibinfo {author} {\bibfnamefont {T.~A.}\ \bibnamefont
  {{Apostolatos}}}, \bibinfo {author} {\bibfnamefont {C.}~\bibnamefont
  {{Cutler}}}, \bibinfo {author} {\bibfnamefont {G.~J.}\ \bibnamefont
  {{Sussman}}},\ and\ \bibinfo {author} {\bibfnamefont {K.~S.}\ \bibnamefont
  {{Thorne}}},\ }\bibfield  {title} {\bibinfo {title} {{Spin-induced orbital
  precession and its modulation of the gravitational waveforms from merging
  binaries}},\ }\href {https://doi.org/10.1103/PhysRevD.49.6274} {\bibfield
  {journal} {\bibinfo  {journal} {\prd}\ }\textbf {\bibinfo {volume} {49}},\
  \bibinfo {pages} {6274} (\bibinfo {year} {1994})}\BibitemShut {NoStop}%
\bibitem [{Note3()}]{Note3}%
  \BibitemOpen
  \bibinfo {note} {Here we have dropped the Shapiro time delay for simplicity,
  which contributes an extra phase $2\pi t_{\protect \rm S}$ with $t_{\protect
  \rm S}=2M_3\protect \qopname \relax o{log}[1/(1-\protect \qopname \relax
  o{sin}\iota _{J}\protect \qopname \relax o{sin}\phi _\protect \text
  {o})]$~\cite {Blandford:76}. For most $\iota _J\protect \neq 90^\circ $,
  $t_{\protect \rm S}\sim 2M_3\ll a_\protect \text {o}$.}\BibitemShut {Stop}%
\bibitem [{\citenamefont {{Dhurandhar}}\ \emph {et~al.}(2005)\citenamefont
  {{Dhurandhar}}, \citenamefont {{Nayak}}, \citenamefont {{Koshti}},\ and\
  \citenamefont {{Vinet}}}]{Dhurandhar:05}%
  \BibitemOpen
  \bibfield  {author} {\bibinfo {author} {\bibfnamefont {S.~V.}\ \bibnamefont
  {{Dhurandhar}}}, \bibinfo {author} {\bibfnamefont {K.~R.}\ \bibnamefont
  {{Nayak}}}, \bibinfo {author} {\bibfnamefont {S.}~\bibnamefont {{Koshti}}},\
  and\ \bibinfo {author} {\bibfnamefont {J.~Y.}\ \bibnamefont {{Vinet}}},\
  }\bibfield  {title} {\bibinfo {title} {{Fundamentals of the LISA stable
  flight formation}},\ }\href {https://doi.org/10.1088/0264-9381/22/3/002}
  {\bibfield  {journal} {\bibinfo  {journal} {Classical and Quantum Gravity}\
  }\textbf {\bibinfo {volume} {22}},\ \bibinfo {pages} {481} (\bibinfo {year}
  {2005})},\ \Eprint {https://arxiv.org/abs/gr-qc/0410093} {arXiv:gr-qc/0410093
  [astro-ph]} \BibitemShut {NoStop}%
\bibitem [{Note4()}]{Note4}%
  \BibitemOpen
  \bibinfo {note} {As a caveat, we note the Fisher matrix may be inaccurate
  when the SNR is low~\cite {Vallisneri:08}. Therefore, future studies consider
  this problem in a full Bayesian framework would be of great
  value.}\BibitemShut {Stop}%
\bibitem [{Note5()}]{Note5}%
  \BibitemOpen
  \bibinfo {note} {See sec.~IV of Ref.~\cite {BCV} for a treatment that gives
  the same waveforms without introducing this apparently discontinuous Thomas
  Precession phase.}\BibitemShut {Stop}%
\bibitem [{\citenamefont {{Heggie}}(1975)}]{Heggie:75}%
  \BibitemOpen
  \bibfield  {author} {\bibinfo {author} {\bibfnamefont {D.~C.}\ \bibnamefont
  {{Heggie}}},\ }\bibfield  {title} {\bibinfo {title} {{Binary evolution in
  stellar dynamics.}},\ }\href {https://doi.org/10.1093/mnras/173.3.729}
  {\bibfield  {journal} {\bibinfo  {journal} {\mnras}\ }\textbf {\bibinfo
  {volume} {173}},\ \bibinfo {pages} {729} (\bibinfo {year}
  {1975})}\BibitemShut {NoStop}%
\bibitem [{\citenamefont {{Barker}}\ and\ \citenamefont
  {{O'Connell}}(1975)}]{Barker:75}%
  \BibitemOpen
  \bibfield  {author} {\bibinfo {author} {\bibfnamefont {B.~M.}\ \bibnamefont
  {{Barker}}}\ and\ \bibinfo {author} {\bibfnamefont {R.~F.}\ \bibnamefont
  {{O'Connell}}},\ }\bibfield  {title} {\bibinfo {title} {{Gravitational
  two-body problem with arbitrary masses, spins, and quadrupole moments}},\
  }\href {https://doi.org/10.1103/PhysRevD.12.329} {\bibfield  {journal}
  {\bibinfo  {journal} {\prd}\ }\textbf {\bibinfo {volume} {12}},\ \bibinfo
  {pages} {329} (\bibinfo {year} {1975})}\BibitemShut {NoStop}%
\bibitem [{\citenamefont {{Barack}}\ and\ \citenamefont
  {{Cutler}}(2004)}]{Barack:04}%
  \BibitemOpen
  \bibfield  {author} {\bibinfo {author} {\bibfnamefont {L.}~\bibnamefont
  {{Barack}}}\ and\ \bibinfo {author} {\bibfnamefont {C.}~\bibnamefont
  {{Cutler}}},\ }\bibfield  {title} {\bibinfo {title} {{LISA capture sources:
  Approximate waveforms, signal-to-noise ratios, and parameter estimation
  accuracy}},\ }\href {https://doi.org/10.1103/PhysRevD.69.082005} {\bibfield
  {journal} {\bibinfo  {journal} {\prd}\ }\textbf {\bibinfo {volume} {69}},\
  \bibinfo {eid} {082005} (\bibinfo {year} {2004})},\ \Eprint
  {https://arxiv.org/abs/gr-qc/0310125} {arXiv:gr-qc/0310125 [gr-qc]}
  \BibitemShut {NoStop}%
\bibitem [{\citenamefont {{Tagawa}}\ \emph {et~al.}(2020)\citenamefont
  {{Tagawa}}, \citenamefont {{Haiman}}, \citenamefont {{Bartos}},\ and\
  \citenamefont {{Kocsis}}}]{Tagawa:20}%
  \BibitemOpen
  \bibfield  {author} {\bibinfo {author} {\bibfnamefont {H.}~\bibnamefont
  {{Tagawa}}}, \bibinfo {author} {\bibfnamefont {Z.}~\bibnamefont {{Haiman}}},
  \bibinfo {author} {\bibfnamefont {I.}~\bibnamefont {{Bartos}}},\ and\
  \bibinfo {author} {\bibfnamefont {B.}~\bibnamefont {{Kocsis}}},\ }\bibfield
  {title} {\bibinfo {title} {{Spin Evolution of Stellar-mass Black Hole
  Binaries in Active Galactic Nuclei}},\ }\href@noop {} {\bibfield  {journal}
  {\bibinfo  {journal} {arXiv e-prints}\ ,\ \bibinfo {eid} {arXiv:2004.11914}}
  (\bibinfo {year} {2020})},\ \Eprint {https://arxiv.org/abs/2004.11914}
  {arXiv:2004.11914 [astro-ph.HE]} \BibitemShut {NoStop}%
\bibitem [{\citenamefont {{Kocsis}}(2013)}]{Kocsis:13}%
  \BibitemOpen
  \bibfield  {author} {\bibinfo {author} {\bibfnamefont {B.}~\bibnamefont
  {{Kocsis}}},\ }\bibfield  {title} {\bibinfo {title} {{High-frequency
  Gravitational Waves from Supermassive Black Holes: Prospects for LIGO-VIRGO
  Detections}},\ }\href {https://doi.org/10.1088/0004-637X/763/2/122}
  {\bibfield  {journal} {\bibinfo  {journal} {\apj}\ }\textbf {\bibinfo
  {volume} {763}},\ \bibinfo {eid} {122} (\bibinfo {year} {2013})},\ \Eprint
  {https://arxiv.org/abs/1211.6427} {arXiv:1211.6427 [astro-ph.HE]}
  \BibitemShut {NoStop}%
\bibitem [{\citenamefont {{Chen}}\ \emph {et~al.}(2019)\citenamefont {{Chen}},
  \citenamefont {{Li}},\ and\ \citenamefont {{Cao}}}]{Chen:19}%
  \BibitemOpen
  \bibfield  {author} {\bibinfo {author} {\bibfnamefont {X.}~\bibnamefont
  {{Chen}}}, \bibinfo {author} {\bibfnamefont {S.}~\bibnamefont {{Li}}},\ and\
  \bibinfo {author} {\bibfnamefont {Z.}~\bibnamefont {{Cao}}},\ }\bibfield
  {title} {\bibinfo {title} {{Mass-redshift degeneracy for the
  gravitational-wave sources in the vicinity of supermassive black holes}},\
  }\href {https://doi.org/10.1093/mnrasl/slz046} {\bibfield  {journal}
  {\bibinfo  {journal} {\mnras}\ }\textbf {\bibinfo {volume} {485}},\ \bibinfo
  {pages} {L141} (\bibinfo {year} {2019})},\ \Eprint
  {https://arxiv.org/abs/1703.10543} {arXiv:1703.10543 [astro-ph.HE]}
  \BibitemShut {NoStop}%
\bibitem [{\citenamefont {{Chen}}(2020)}]{Chen:20b}%
  \BibitemOpen
  \bibfield  {author} {\bibinfo {author} {\bibfnamefont {X.}~\bibnamefont
  {{Chen}}},\ }\bibfield  {title} {\bibinfo {title} {{Distortion of
  Gravitational-wave Signals by Astrophysical Environments}},\ }\href@noop {}
  {\bibfield  {journal} {\bibinfo  {journal} {arXiv e-prints}\ ,\ \bibinfo
  {eid} {arXiv:2009.07626}} (\bibinfo {year} {2020})},\ \Eprint
  {https://arxiv.org/abs/2009.07626} {arXiv:2009.07626 [astro-ph.HE]}
  \BibitemShut {NoStop}%
\bibitem [{\citenamefont {{Blandford}}\ and\ \citenamefont
  {{Teukolsky}}(1976)}]{Blandford:76}%
  \BibitemOpen
  \bibfield  {author} {\bibinfo {author} {\bibfnamefont {R.}~\bibnamefont
  {{Blandford}}}\ and\ \bibinfo {author} {\bibfnamefont {S.~A.}\ \bibnamefont
  {{Teukolsky}}},\ }\bibfield  {title} {\bibinfo {title} {{Arrival-time
  analysis for a pulsar in a binary system.}},\ }\href
  {https://doi.org/10.1086/154315} {\bibfield  {journal} {\bibinfo  {journal}
  {\apj}\ }\textbf {\bibinfo {volume} {205}},\ \bibinfo {pages} {580} (\bibinfo
  {year} {1976})}\BibitemShut {NoStop}%
\bibitem [{\citenamefont {{Vallisneri}}(2008)}]{Vallisneri:08}%
  \BibitemOpen
  \bibfield  {author} {\bibinfo {author} {\bibfnamefont {M.}~\bibnamefont
  {{Vallisneri}}},\ }\bibfield  {title} {\bibinfo {title} {{Use and abuse of
  the Fisher information matrix in the assessment of gravitational-wave
  parameter-estimation prospects}},\ }\href
  {https://doi.org/10.1103/PhysRevD.77.042001} {\bibfield  {journal} {\bibinfo
  {journal} {\prd}\ }\textbf {\bibinfo {volume} {77}},\ \bibinfo {eid} {042001}
  (\bibinfo {year} {2008})},\ \Eprint {https://arxiv.org/abs/gr-qc/0703086}
  {arXiv:gr-qc/0703086 [gr-qc]} \BibitemShut {NoStop}%
\bibitem [{\citenamefont {Buonanno}\ \emph {et~al.}(2003)\citenamefont
  {Buonanno}, \citenamefont {Chen},\ and\ \citenamefont {Vallisneri}}]{BCV}%
  \BibitemOpen
  \bibfield  {author} {\bibinfo {author} {\bibfnamefont {A.}~\bibnamefont
  {Buonanno}}, \bibinfo {author} {\bibfnamefont {Y.}~\bibnamefont {Chen}},\
  and\ \bibinfo {author} {\bibfnamefont {M.}~\bibnamefont {Vallisneri}},\
  }\bibfield  {title} {\bibinfo {title} {Detecting gravitational waves from
  precessing binaries of spinning compact objects: Adiabatic limit},\
  }\href@noop {} {\bibfield  {journal} {\bibinfo  {journal} {Physical Review
  D}\ }\textbf {\bibinfo {volume} {67}},\ \bibinfo {pages} {104025} (\bibinfo
  {year} {2003})}\BibitemShut {NoStop}%
\end{thebibliography}%


\begin{thebibliography}{12}%
\makeatletter
\providecommand \@ifxundefined [1]{%
 \@ifx{#1\undefined}
}%
\providecommand \@ifnum [1]{%
 \ifnum #1\expandafter \@firstoftwo
 \else \expandafter \@secondoftwo
 \fi
}%
\providecommand \@ifx [1]{%
 \ifx #1\expandafter \@firstoftwo
 \else \expandafter \@secondoftwo
 \fi
}%
\providecommand \natexlab [1]{#1}%
\providecommand \enquote  [1]{``#1''}%
\providecommand \bibnamefont  [1]{#1}%
\providecommand \bibfnamefont [1]{#1}%
\providecommand \citenamefont [1]{#1}%
\providecommand \href@noop [0]{\@secondoftwo}%
\providecommand \href [0]{\begingroup \@sanitize@url \@href}%
\providecommand \@href[1]{\@@startlink{#1}\@@href}%
\providecommand \@@href[1]{\endgroup#1\@@endlink}%
\providecommand \@sanitize@url [0]{\catcode `\\12\catcode `\$12\catcode
  `\&12\catcode `\#12\catcode `\^12\catcode `\_12\catcode `\%12\relax}%
\providecommand \@@startlink[1]{}%
\providecommand \@@endlink[0]{}%
\providecommand \url  [0]{\begingroup\@sanitize@url \@url }%
\providecommand \@url [1]{\endgroup\@href {#1}{\urlprefix }}%
\providecommand \urlprefix  [0]{URL }%
\providecommand \Eprint [0]{\href }%
\providecommand \doibase [0]{https://doi.org/}%
\providecommand \selectlanguage [0]{\@gobble}%
\providecommand \bibinfo  [0]{\@secondoftwo}%
\providecommand \bibfield  [0]{\@secondoftwo}%
\providecommand \translation [1]{[#1]}%
\providecommand \BibitemOpen [0]{}%
\providecommand \bibitemStop [0]{}%
\providecommand \bibitemNoStop [0]{.\EOS\space}%
\providecommand \EOS [0]{\spacefactor3000\relax}%
\providecommand \BibitemShut  [1]{\csname bibitem#1\endcsname}%
\let\auto@bib@innerbib\@empty
\bibitem [{\citenamefont {{Liu}}\ \emph {et~al.}(2019)\citenamefont {{Liu}},
  \citenamefont {{Lai}},\ and\ \citenamefont {{Wang}}}]{Liu:19}%
  \BibitemOpen
  \bibfield  {author} {\bibinfo {author} {\bibfnamefont {B.}~\bibnamefont
  {{Liu}}}, \bibinfo {author} {\bibfnamefont {D.}~\bibnamefont {{Lai}}},\ and\
  \bibinfo {author} {\bibfnamefont {Y.-H.}\ \bibnamefont {{Wang}}},\ }\bibfield
   {title} {\bibinfo {title} {{Binary Mergers near a Supermassive Black Hole:
  Relativistic Effects in Triples}},\ }\href
  {https://doi.org/10.3847/2041-8213/ab40c0} {\bibfield  {journal} {\bibinfo
  {journal} {\apjl}\ }\textbf {\bibinfo {volume} {883}},\ \bibinfo {eid} {L7}
  (\bibinfo {year} {2019})},\ \Eprint {https://arxiv.org/abs/1906.07726}
  {arXiv:1906.07726 [astro-ph.HE]} \BibitemShut {NoStop}%
\bibitem [{\citenamefont {{Yu}}\ \emph {et~al.}(2020)\citenamefont {{Yu}},
  \citenamefont {{Ma}}, \citenamefont {{Giesler}},\ and\ \citenamefont
  {{Chen}}}]{Yu:20b}%
  \BibitemOpen
  \bibfield  {author} {\bibinfo {author} {\bibfnamefont {H.}~\bibnamefont
  {{Yu}}}, \bibinfo {author} {\bibfnamefont {S.}~\bibnamefont {{Ma}}}, \bibinfo
  {author} {\bibfnamefont {M.}~\bibnamefont {{Giesler}}},\ and\ \bibinfo
  {author} {\bibfnamefont {Y.}~\bibnamefont {{Chen}}},\ }\bibfield  {title}
  {\bibinfo {title} {{Spin and eccentricity evolution in triple systems: From
  the Lidov-Kozai interaction to the final merger of the inner binary}},\
  }\href {https://doi.org/10.1103/PhysRevD.102.123009} {\bibfield  {journal}
  {\bibinfo  {journal} {\prd}\ }\textbf {\bibinfo {volume} {102}},\ \bibinfo
  {eid} {123009} (\bibinfo {year} {2020})},\ \Eprint
  {https://arxiv.org/abs/2007.12978} {arXiv:2007.12978 [gr-qc]} \BibitemShut
  {NoStop}%
\bibitem [{\citenamefont {{Ostriker}}(1999)}]{Ostriker:99}%
  \BibitemOpen
  \bibfield  {author} {\bibinfo {author} {\bibfnamefont {E.~C.}\ \bibnamefont
  {{Ostriker}}},\ }\bibfield  {title} {\bibinfo {title} {{Dynamical Friction in
  a Gaseous Medium}},\ }\href {https://doi.org/10.1086/306858} {\bibfield
  {journal} {\bibinfo  {journal} {\apj}\ }\textbf {\bibinfo {volume} {513}},\
  \bibinfo {pages} {252} (\bibinfo {year} {1999})},\ \Eprint
  {https://arxiv.org/abs/astro-ph/9810324} {arXiv:astro-ph/9810324 [astro-ph]}
  \BibitemShut {NoStop}%
\bibitem [{\citenamefont {{Chen}}\ \emph {et~al.}(2020)\citenamefont {{Chen}},
  \citenamefont {{Xuan}},\ and\ \citenamefont {{Peng}}}]{Chen:20}%
  \BibitemOpen
  \bibfield  {author} {\bibinfo {author} {\bibfnamefont {X.}~\bibnamefont
  {{Chen}}}, \bibinfo {author} {\bibfnamefont {Z.-Y.}\ \bibnamefont {{Xuan}}},\
  and\ \bibinfo {author} {\bibfnamefont {P.}~\bibnamefont {{Peng}}},\
  }\bibfield  {title} {\bibinfo {title} {{Fake Massive Black Holes in the
  Milli-Hertz Gravitational-wave Band}},\ }\href
  {https://doi.org/10.3847/1538-4357/ab919f} {\bibfield  {journal} {\bibinfo
  {journal} {\apj}\ }\textbf {\bibinfo {volume} {896}},\ \bibinfo {eid} {171}
  (\bibinfo {year} {2020})},\ \Eprint {https://arxiv.org/abs/2003.08639}
  {arXiv:2003.08639 [astro-ph.HE]} \BibitemShut {NoStop}%
\bibitem [{\citenamefont {{Antoni}}\ \emph {et~al.}(2019)\citenamefont
  {{Antoni}}, \citenamefont {{MacLeod}},\ and\ \citenamefont
  {{Ramirez-Ruiz}}}]{Antoni:19}%
  \BibitemOpen
  \bibfield  {author} {\bibinfo {author} {\bibfnamefont {A.}~\bibnamefont
  {{Antoni}}}, \bibinfo {author} {\bibfnamefont {M.}~\bibnamefont
  {{MacLeod}}},\ and\ \bibinfo {author} {\bibfnamefont {E.}~\bibnamefont
  {{Ramirez-Ruiz}}},\ }\bibfield  {title} {\bibinfo {title} {{The Evolution of
  Binaries in a Gaseous Medium: Three-dimensional Simulations of Binary
  Bondi-Hoyle-Lyttleton Accretion}},\ }\href
  {https://doi.org/10.3847/1538-4357/ab3466} {\bibfield  {journal} {\bibinfo
  {journal} {\apj}\ }\textbf {\bibinfo {volume} {884}},\ \bibinfo {eid} {22}
  (\bibinfo {year} {2019})},\ \Eprint {https://arxiv.org/abs/1901.07572}
  {arXiv:1901.07572 [astro-ph.HE]} \BibitemShut {NoStop}%
\bibitem [{\citenamefont {{Baruteau}}\ \emph {et~al.}(2011)\citenamefont
  {{Baruteau}}, \citenamefont {{Cuadra}},\ and\ \citenamefont
  {{Lin}}}]{Baruteau:11}%
  \BibitemOpen
  \bibfield  {author} {\bibinfo {author} {\bibfnamefont {C.}~\bibnamefont
  {{Baruteau}}}, \bibinfo {author} {\bibfnamefont {J.}~\bibnamefont
  {{Cuadra}}},\ and\ \bibinfo {author} {\bibfnamefont {D.~N.~C.}\ \bibnamefont
  {{Lin}}},\ }\bibfield  {title} {\bibinfo {title} {{Binaries Migrating in a
  Gaseous Disk: Where are the Galactic Center Binaries?}},\ }\href
  {https://doi.org/10.1088/0004-637X/726/1/28} {\bibfield  {journal} {\bibinfo
  {journal} {\apj}\ }\textbf {\bibinfo {volume} {726}},\ \bibinfo {eid} {28}
  (\bibinfo {year} {2011})},\ \Eprint {https://arxiv.org/abs/1011.0360}
  {arXiv:1011.0360 [astro-ph.GA]} \BibitemShut {NoStop}%
\bibitem [{\citenamefont {{Bartos}}\ \emph {et~al.}(2017)\citenamefont
  {{Bartos}}, \citenamefont {{Kocsis}}, \citenamefont {{Haiman}},\ and\
  \citenamefont {{M{\'a}rka}}}]{Bartos:17}%
  \BibitemOpen
  \bibfield  {author} {\bibinfo {author} {\bibfnamefont {I.}~\bibnamefont
  {{Bartos}}}, \bibinfo {author} {\bibfnamefont {B.}~\bibnamefont {{Kocsis}}},
  \bibinfo {author} {\bibfnamefont {Z.}~\bibnamefont {{Haiman}}},\ and\
  \bibinfo {author} {\bibfnamefont {S.}~\bibnamefont {{M{\'a}rka}}},\
  }\bibfield  {title} {\bibinfo {title} {{Rapid and Bright Stellar-mass Binary
  Black Hole Mergers in Active Galactic Nuclei}},\ }\href
  {https://doi.org/10.3847/1538-4357/835/2/165} {\bibfield  {journal} {\bibinfo
   {journal} {\apj}\ }\textbf {\bibinfo {volume} {835}},\ \bibinfo {eid} {165}
  (\bibinfo {year} {2017})},\ \Eprint {https://arxiv.org/abs/1602.03831}
  {arXiv:1602.03831 [astro-ph.HE]} \BibitemShut {NoStop}%
\bibitem [{\citenamefont {{G{\"u}ltekin}}\ \emph {et~al.}(2004)\citenamefont
  {{G{\"u}ltekin}}, \citenamefont {{Miller}},\ and\ \citenamefont
  {{Hamilton}}}]{Gultekin:04}%
  \BibitemOpen
  \bibfield  {author} {\bibinfo {author} {\bibfnamefont {K.}~\bibnamefont
  {{G{\"u}ltekin}}}, \bibinfo {author} {\bibfnamefont {M.~C.}\ \bibnamefont
  {{Miller}}},\ and\ \bibinfo {author} {\bibfnamefont {D.~P.}\ \bibnamefont
  {{Hamilton}}},\ }\bibfield  {title} {\bibinfo {title} {{Growth of
  Intermediate-Mass Black Holes in Globular Clusters}},\ }\href
  {https://doi.org/10.1086/424809} {\bibfield  {journal} {\bibinfo  {journal}
  {\apj}\ }\textbf {\bibinfo {volume} {616}},\ \bibinfo {pages} {221} (\bibinfo
  {year} {2004})},\ \Eprint {https://arxiv.org/abs/astro-ph/0402532}
  {arXiv:astro-ph/0402532 [astro-ph]} \BibitemShut {NoStop}%
\bibitem [{\citenamefont {{Antonini}}\ and\ \citenamefont
  {{Rasio}}(2016)}]{Antonini:16}%
  \BibitemOpen
  \bibfield  {author} {\bibinfo {author} {\bibfnamefont {F.}~\bibnamefont
  {{Antonini}}}\ and\ \bibinfo {author} {\bibfnamefont {F.~A.}\ \bibnamefont
  {{Rasio}}},\ }\bibfield  {title} {\bibinfo {title} {{Merging Black Hole
  Binaries in Galactic Nuclei: Implications for Advanced-LIGO Detections}},\
  }\href {https://doi.org/10.3847/0004-637X/831/2/187} {\bibfield  {journal}
  {\bibinfo  {journal} {\apj}\ }\textbf {\bibinfo {volume} {831}},\ \bibinfo
  {eid} {187} (\bibinfo {year} {2016})},\ \Eprint
  {https://arxiv.org/abs/1606.04889} {arXiv:1606.04889 [astro-ph.HE]}
  \BibitemShut {NoStop}%
\bibitem [{Note1()}]{Note1}%
  \BibitemOpen
  \bibinfo {note} {Here we treat $\gamma $ as a constant for simplicity and
  ignore the general-relativistic precession of the pericenter which happens at
  a rate $\Omega _{e}\simeq 3M_3 \Omega _{\protect \text {o}}/\left [a_\protect
  \text {o}(1-e_\protect \text {o}^2)\right ]\simeq 2\Omega _{\protect \rm
  dS}$. Including it would introduce more dynamics on the system without
  needing extra unknown parameters. Consequently it would make the constraints
  on $(M_3, a_\protect \text {o}, e_\protect \text {o})$ even
  better.}\BibitemShut {Stop}%
\bibitem [{\citenamefont {{Flanagan}}\ and\ \citenamefont
  {{Hughes}}(1998)}]{Flanagan:98}%
  \BibitemOpen
  \bibfield  {author} {\bibinfo {author} {\bibfnamefont {{\'E}.~{\'E}.}\
  \bibnamefont {{Flanagan}}}\ and\ \bibinfo {author} {\bibfnamefont {S.~A.}\
  \bibnamefont {{Hughes}}},\ }\bibfield  {title} {\bibinfo {title} {{Measuring
  gravitational waves from binary black hole coalescences. I. Signal to noise
  for inspiral, merger, and ringdown}},\ }\href
  {https://doi.org/10.1103/PhysRevD.57.4535} {\bibfield  {journal} {\bibinfo
  {journal} {\prd}\ }\textbf {\bibinfo {volume} {57}},\ \bibinfo {pages} {4535}
  (\bibinfo {year} {1998})},\ \Eprint {https://arxiv.org/abs/gr-qc/9701039}
  {arXiv:gr-qc/9701039 [gr-qc]} \BibitemShut {NoStop}%
\bibitem [{\citenamefont {{Barack}}\ and\ \citenamefont
  {{Cutler}}(2004)}]{Barack:04}%
  \BibitemOpen
  \bibfield  {author} {\bibinfo {author} {\bibfnamefont {L.}~\bibnamefont
  {{Barack}}}\ and\ \bibinfo {author} {\bibfnamefont {C.}~\bibnamefont
  {{Cutler}}},\ }\bibfield  {title} {\bibinfo {title} {{LISA capture sources:
  Approximate waveforms, signal-to-noise ratios, and parameter estimation
  accuracy}},\ }\href {https://doi.org/10.1103/PhysRevD.69.082005} {\bibfield
  {journal} {\bibinfo  {journal} {\prd}\ }\textbf {\bibinfo {volume} {69}},\
  \bibinfo {eid} {082005} (\bibinfo {year} {2004})},\ \Eprint
  {https://arxiv.org/abs/gr-qc/0310125} {arXiv:gr-qc/0310125 [gr-qc]}
  \BibitemShut {NoStop}%
\end{thebibliography}%

\end{document}




\title{Supplemental material for ``Direct determination of supermassive black hole properties with gravitational-wave radiation from surrounding stellar-mass black hole binaries''}

\maketitle
\section{Explicit expressions for various timescales}
Here we provide explicit expressions of various relevant timescales.

The instantaneous GW decay timescale is
\begin{align}
    \tau_{\rm gw}&=\frac{a}{|\dot{a}|} = \frac{5}{64}\frac{a^4}{\mu M_{\rm t}^2}\left[\frac{\left(1-e^2\right)^{7/2}}{1+\frac{73}{24}e^2 + \frac{37}{96}e^4}\right],\nonumber \\
    &=20\,{\rm yr} \left[\frac{\left(1-e^2\right)^{7/2}}{1+\frac{73}{24}e^2 + \frac{37}{96}e^4}\right]\left(\frac{\mathcal{M}}{44\,M_\odot}\right)^{-5/3}\left(\frac{2f_{\rm orb}}{12\,{\rm mHz}}\right)^{-8/3},
\end{align}
where $\mu$, $M_{\rm t}$, and $\mathcal{M}$ are respectively the reduced mass, total mass, and the chirp mass of the binary of interest. In the second line we have scaled the number by the orbital frequency for future convenience, though we remind the reader that the timescale defined here is the instantaneous decay rate of the semi-major axis (instead of frequency). Furthermore, when the orbit is circular, the GW radiation has a single frequency component with $f=2f_{\rm orb}$, and thus a factor of 2 is included in the scaling of $f_{\rm orb}$. 

For a circular orbit, the total time to merger is $t_{\rm m}=\tau_{\rm gw}/4$. By setting $t_{\rm m}=T_{\rm obs}=5\,{\rm yr}$ (the fiducial observation time), we can then determine the initial frequency (or the initial orbital separation) for a given binary system. This is why we chose an initial GW frequency  $f^{(0)}=2f_{\rm orb}=12\,{\rm mHz}$ ($a_{\imag}^{(0)}=1.4\times10^{-3}\,{\rm AU}$) for the binary with $M_1=M_2=50\,{M_\odot}$ and $f^{(0)}=2f_{\rm orb}=4.4\,{\rm mHz}$ ($a_{\imag}^{(0)}=4.7\times10^{-3}$) for the one with $M_1=M_2=250\,M_\odot$ (see Fig. 3 in the main text). In comparison, for a typical outer orbit with $M_1+M_2=100\,M_\odot$, $M_3=10^8\,M_\odot$, and $a_\out=100 M_3\simeq 100\,{\rm AU}$, the merger time is $t_{\rm m,\out}\simeq 3\times 10^{7}\,{\rm yr}$. Therefore, in most cases we can safely ignore the GW-induced decay of the outer orbit. 

The presence of the central SMBH will modulate the GW waveform emitted by the inner binary (i.e., the carrier) via various effects. The most significant one is the Doppler phase shift due to the motion of the outer orbit (Newtonian dipole effect), at a rate $\Omega_\out\simeq \sqrt{M_3/a_\out^3}$, or a period
\begin{equation}
    P_\out=\frac{2\pi}{\Omega_\out} = 0.1\,{\rm yr}\left(\frac{M_3}{10^8\,M_\odot}\right)\left(\frac{a_\out}{100 M_3}\right)^{3/2}. 
\end{equation}

The next leading-order effect is the de Sitter-like precession of the inner orbit (a 1.5 post-Newtonian-order, or 1.5 PN effect), which is the focus of the main text. It has a rate~\cite{Liu:19, Yu:20b}
\begin{equation}
    \Omega_{\rm dS} = \frac{3}{2}\frac{M_3+\mu_\out/3}{a_\out(1-e_\out^2)}\Omega_\out\simeq \frac{3}{2}\frac{M_3}{a_\out(1-e_\out^2)}\Omega_\out,
\end{equation}
where the second equality applies because $M_3\gg \mu_\out\simeq (M_1+M_2)$. The corresponding period is thus
\begin{equation}
    P_{\rm dS}=6.5\,{\rm yr}\left(1-e_\out^2\right)\left(\frac{M_3}{10^8\,M_\odot}\right)\left(\frac{a_\out}{100 M_3}\right)^{5/2}. 
\end{equation}

When the central SMBH is fast spinning, the inner orbit will also precess around the spin of the SMBH $\vect{S}_3$ by the Lense-Thirring effect (2 PN). Its rate is
\begin{equation}
    \Omega_{\rm LT} = \frac{S_3}{a_\out^3(1-e_\out)^{3/2}},
\end{equation}
and period 
\begin{equation}
    P_{\rm LT} = 2.0\times10^2\,{\rm yr}(1-e_\out^2)^{3/2}\left(\frac{S_3}{M_3^2}\right)\left(\frac{M_3}{10^8\,M_\odot}\right)\left(\frac{a_\out}{100 M_3}\right)^{3}. 
\end{equation}

Additionally, the SMBH may perturb the inner orbit via the Lidov-Kozai effect (i.e., the Newtonian tidal effect as it comes at the quadrupole order). The rate is given by~\cite{Liu:19}
\begin{equation}
    \Omega_{\rm LK} = \frac{M_3}{(M_1+M_2)}\left(\frac{a_\imag}{a_\out \sqrt{1-e_\out^2}}\right)^3\Omega_i,
\end{equation}
where $\Omega_\imag=\sqrt{(M_1+M_2)/a_\imag^3}$ is the orbital frequency of the inner orbit. 
The corresponding period is thus 
\begin{equation}
    P_{\rm LK} =1.8\times10^{3}\,{\rm yr}(1-e_\out^2)^{3/2}
    \left(\frac{M_3}{10^8\,M_\odot}\right)^2
    \left(\frac{a_\out}{100 M_3}\right)^{3} 
    \left(\frac{M_1+M_2}{100\,M_\odot}\right)^{1/2}
    \left(\frac{a_\imag}{1.4\times10^{-3}\,{\rm AU}}\right)^{-3/2}.
\end{equation}
Unlike the de Sitter and Lense-Thirring effects which are independent of $a_\imag$, the Lidov-Kozai timescale increases as the inner orbit decays because the ``lever arm'' for the SMBH to perturb is smaller. The Lidov-Kozai effect is therefore less and less significant as the inner binary evolves towards the merger.

As the inner binary may reside in a gaseous disk, the frictional force from the background gas may both cause the inner binary as a whole to accelerate/decelerate from the Keplerain outer orbit, and harden the inner binary and make it merges in a shorter timescale than $t_{\rm m}$. 

For the gaseous effect on the outer orbit, we estimate it with the dynamical friction derived in Ref.~\cite{Ostriker:99}, which leads to a characteristic timescale~\cite{Chen:20}
\begin{equation}
    \tau_{\rm gas}\equiv\frac{a_\out}{|\dot{a}_{\rm o, gas}|}=8\times10^{5}\,{\rm yr}\left(\frac{\rho_{\rm bg}}{10^{-8}\,{\rm g\,cm^{-3}}}\right)^{-1}\left(\frac{M_1+M_2}{100\,M_\odot}\right)^{-1}\left(\frac{a_\out}{100 M_3}\right)^{-3/2},
\end{equation}
where $\dot{a}_{\out, {\rm gas}}$ is the rate at which the outer orbit changes due to the hydrodynamic drag and $\rho_{\rm bg}$ is the background gas density. Although this effect may be important for the migration of the outer orbit over the entire evolution of the inner binary, over a period of $T_{\rm obs}\simeq 5\,{\rm yr}$, it only changes the outer orbit by a fractional amount of $T_{\rm obs}/\tau_{\rm gas}\sim 10^{-5}$ and can thus be safely ignored.

As for the inner binary, a circumbinary mini-accretion disk may form. In this scenario, the inner binary hardens due to the gaseous effect over a timescale~\cite{Chen:20}
\begin{equation}
    \tau'_{\rm gas}=4\times10^{3}\,{\rm yr}\times q^{-1}\left(\frac{2}{1+q}\right)^{-3}\left(\frac{M_1}{50\,{\rm M_\odot}}\right)^{-1}\left(\frac{\rho_{\rm bg}}{10^{-13}\,{\rm g\,cm^{-3}}}\right)^{-1}\left(\frac{c_{\rm s}}{10^2\,{\rm km\,s^{-1}}}\right)^3,
\end{equation}
where $q=M_2/M_1$. A similar estimation can be found in Ref.~\cite{Antoni:19} where the authors found the inspiraling rate changes from gas dominated to GW dominated at an inner separation of $a_\imag\sim R_\odot\simeq 5\times 10^{-3}\,{\rm AU}$. For the typical $a_\imag$ we consider, this then indicates $\tau'_{\rm gas}\sim 100\tau_{\rm gw}$. 

Ref.~\cite{Baruteau:11} suggested yet another hardening mechanism due to the formation of overdense spiral tails lagging the BHs in the inner binary and exerting torques on them. This mechanism could efficiently half the inner semi-major axis in a few cycles of the outer orbit, and for $a_\out \sim 100 M_3$, such a timescale could be comparable to the duration of observation. Nonetheless, the model considered by Ref.~\cite{Baruteau:11} applies for inner binaries with separations of $a_\imag\sim r_{\rm H}$, where $r_{\rm H} = a_\out \left[M_3/3(M_1+M_2)\right]^{1/3}$ is the Hill radius. For $a_\out=100\,M_3$ and $M_3=10^8\,M_\odot$ ($M_3=10^6\,M_\odot$), we have $r_{\rm H} \simeq 0.7\,{\rm AU}$ ($r_{\rm H} \simeq 0.03\,{\rm AU}$), much greater than the initial inner binary's separation of $a_\imag\simeq 1.4\times10^{-3}\,{\rm AU}$ considered in our work. Therefore, our case is likely to be beyond the regime of validity of the model proposed in~\cite{Baruteau:11} (see also the discussion in sec. 8.5 of Ref.~\cite{Baruteau:11} and sec. 2.3 of Ref.~\cite{Bartos:17}). As a result, this is an effect critical for the early evolution of the inner binary but is likely subdominant for the final state when $\tau_{\rm gw}=\mathcal{O}(10\,{\rm yr})$.

Furthermore, as shown in Ref.~\cite{Chen:20}, the gaseous friction's effect on the inner binary is make the chirp mass appear heavier than the true value by a factor $(1+\tau'_{\rm gas}/\tau_{\rm gw})^{3/5}$. It can therefore be absorbed into the carrier waveform $\tilde{h}_{\rm c}$ and be extracted from the frequency evolution of the waveform similar to high-order post-Newtonian parameter. 

The inner binary, after formation, may also experience multiple encounter with the surrounding background stars/BHs. The typical timescale between two consecutive interactions can be estimated to be~\cite{Gultekin:04, Antonini:16}
\begin{equation}
    \tau_{\rm enc}=2\times 10^{5}\,{\rm yr} \left(\frac{\sigma}{0.01}\right)\left(\frac{n_{\rm bg}}{10^{10}\,{\rm pc^{-3}}}\right)\left(\frac{r_{\rm p}}{0.01\,{\rm AU}}\right)^{-1}\left(\frac{M_1+M_2}{100\,M_\odot}\right)^{-1}\left(\frac{M_{\rm bg}}{10\,M_\odot}\right)^{-1/2},
\end{equation}
where $\sigma$ is the velocity dispersion, $n_{\rm bg}$ the number density of background stars/BHs, $r_{\rm p}$ the maximum considered close approach to the inner binary, and $M_{\rm bg}$ the mass of the background perturber. In the scaling above, we have conservatively (making $\tau_{\rm enc}$ smaller) set $\sigma = 0.1 v_{\rm orb}(a_\out=100\,M_3)$ and $r_{\rm p}\simeq 10 a_\imag$. Note $\tau_{\rm enc}\propto r_{\rm p}^{-1}\sim a_\imag^{-1}$, and therefore encounters with background objects are important when the inner binary is far apart (e.g., when it is just formed). The frequent encounters at the early stages also play a critical role in giving the inner binary a nearly isotropic orientation so that $\vect{L}_\imag$ is typically misaligned with $\vect{L}_\out$. However, at the end stage of the inner binary's evolution with $\tau_{\rm gw}\sim T_{\rm obs}$, we have $\tau_{\rm gw}\ll \tau_{\rm enc}$, and therefore it is very unlikely for the inner binary to be disrupted during the observation. 

\section{Explicit expressions for the waveforms}
Here we provide explicit expressions for various quantities used in our construction of the waveform. 

The ``carrier'' waveform in our study is given by 
\begin{equation}
    \tilde{h}_{\rm C}(f) = \left(\frac{5}{96}\right)^{1/2}\frac{\mathcal{M}^{5/6}}{\pi^{2/3}D_L}f^{-7/6} \exp\left\{\imag \left[2\pi f t_{\rm c} -\phi_{\rm c}-\frac{\pi}{4}+\frac{3}{4}(8\pi\mathcal{M}f)^{-5/3}\right]\right\}.
\end{equation}

The antenna pattern coefficients are 
\begin{align}
    &F_{+}(\theta_{\rm S}, \phi_{\rm S}, \psi_{\rm S}) = \frac{1}{2}\left(1+\cos^2\theta_{\rm S}\right)\cos2\phi_{\rm S}\cos2\psi_{\rm S} - \cos\theta_{\rm S}\sin2\phi_{\rm S}\sin2\psi_{\rm S}, \\
    &F_\times(\theta_{\rm S}, \phi_{\rm S}, \psi_{\rm S})=\frac{1}{2}\left(1+\cos^2\theta_{\rm S}\right)\cos2\phi_{\rm S}\sin2\psi_{\rm S} + \cos\theta_{\rm S}\sin2\phi_{\rm S}\cos2\psi_{\rm S},
\end{align}
where $(\theta_{\rm S}, \phi_{\rm S})$ are the polar coordinates of $\uvect{N}$ in the time-varying $(x,y,z)$ frame, and 
\begin{equation}
    \psi_{\rm S} = \tan^{-1}\left[\uvect{L}_\imag\cdot\uvect{z} - \frac{ (\uvect{L}_\imag\cdot\uvect{N})(\uvect{z}\cdot\uvect{N})}{\uvect{N}\cdot(\uvect{L}_\imag\times \uvect{z})}\right]
\end{equation}
is the polarization angle of the source. 

We calculate the Thomas phase $\Phi_{\rm T}$ by integrating
\begin{equation}
    \Phi_{\rm T}(t) = -\int_t^{t_{\rm c}} dt \left[\frac{\uvect{L}\cdot\uvect{N}}{1-\left(\uvect{L}\cdot\uvect{N}\right)^2}\right]\left(\uvect{L}\times \uvect{N}\right)\cdot \frac{d\uvect{L}}{dt},\label{eq:Phi_T}
\end{equation}
and the polarization phase $\Phi_{\rm P}$ from the relation
\begin{equation}
    \Phi_{\rm P}(t) = \arctan \left[\frac{-A_\times(t) F_\times(t)}{A_+(t) F_+(t)}\right].
\end{equation}

The time-dependent orientation of $\uvect{L}_\imag$ in our case is given by
\begin{align}
    \uvect{L}_\imag&= \left[\cos\lambda_L \sin\overline{\theta}_J\cos\overline{\phi}_J
    +\sin\lambda_L\left(-\cos\overline{\theta}_J \cos\overline{\phi}_J\cos\alpha+ \sin\overline{\phi}_J\sin \alpha \right)\right]\uvect{\overline{x}}  \nonumber \\ 
    & + \left[\cos\lambda_L\sin\overline{\theta}_J\sin\overline{\phi}_J 
    -\sin\lambda_L\left(\cos\overline{\phi}_J\sin\alpha + \cos\overline{\theta}_J\sin\overline{\phi}_J\cos\alpha\right)\right]\uvect{\overline{y}} \nonumber \\
    &+\left[\cos\lambda_L\cos\overline{\theta}_J+\sin\lambda_L\sin\overline{\theta}_J\cos\alpha\right]\uvect{\overline{z}},
\end{align}
where $\alpha = \Omega_{\rm dS} t + \alpha_0$. 

The detector's orientations are
\begin{align}
    &\uvect{z}(t)= - \frac{\sqrt{3}}{2}\left(\cos\phi_{\rm d}\uvect{\overline{x}} + \sin\phi_{\rm d}\uvect{\overline{y}} \right) + \frac{1}{2}\uvect{\overline{z}}. \\
    &\uvect{x}(t) = -\frac{\sin2\phi_{\rm d}}{4}\uvect{\overline{x}} 
    + \frac{\left.3+\cos2\phi_{\rm d}\right.}{4}\uvect{\overline{y}}
    + \frac{\sqrt{3}}{2} \sin\phi_{\rm d}\uvect{\overline{z}}
\end{align}
and $\uvect{y} = \uvect{z}\times \uvect{x}$. In the expressions above, $\phi_{\rm d}=2\pi t/{\rm yr}$ is the phase of the detector. 

To summarize, when we consider the simple-precession problem, the waveform is parameterized in terms 11 free parameters in total, $(\mathcal{M}, D_{\rm L}, t_{\rm c}, \phi_c, \overline{\theta}_{\rm S}, \overline{\phi}_{\rm S}, \overline{\theta}_J, \overline{\phi}_J, P_{\rm dS}, \lambda_L, \alpha_0)$. When consider the full SMBH effects (dS precession and Doppler shift due to the outer orbital motion), we further write $P_{\rm dS}$ in terms of $M_3$ and $a_\out$, and include $\phi^{(0)}$ as the initial phase of the outer orbit's Doppler phase.

\section{Doppler phase shift of elliptic outer orbits}
Here we demonstrate that we can extract simultaneously the orbital period (hence the enclosed mass density) and eccentricity of an elliptic outer orbit from the Doppler phase shift alone. 

To do so, we consider a simple model with $\tilde{h}(f)=\tilde{h}_{\rm C}(f)\exp[-i\Phi_{\rm D}(t)]$. 
In other words, we include only the Doppler phase shift due to the outer orbit (now has finite eccentricity) and drop other antenna responses for simplicity. The Doppler phase can be further written as 
\begin{equation}
    \Phi_{\rm D}(t) = 2\pi f r_{\rm o, \parallel}(t),
\end{equation}
where $r_{\rm o, \parallel}(t)$ is the orbital separation projected along the line of sight. Specifically, we have
\begin{equation}
    r_{\rm o, \parallel}(t) = \frac{\mathcal{A}(1-e_\out^2)}{1+e_\out\cos u(t)}\sin[u(t)+\gamma],
\end{equation}
where $u(t)$ and $\gamma$ are the true anomoly and the argument of pericenter.\footnote{Here we treat $\gamma$ as a constant for simplicity and ignore the general-relativistic precession of the pericenter which happens at a rate $\Omega_{e}\simeq 3M_3 \Omega_{\out}/\left[a_\out (1-e_\out^2)\right]\simeq 2\Omega_{\rm dS}$. Including it would introduce more dynamics on the system without needing extra unknown parameters. Consequently it would make the constraints on $(M_3, a_\out, e_\out)$ even better. } The amplitude is further given by $\mathcal{A} = a_\out\sin \iota$. Note that with Doppler shift alone we cannot separate out $a_\out$ and $\sin \iota$, and thus we treat  $\mathcal{A}$ itself as a free parameter. 
The true anomoly can be solved as a function of time (which is further a fucntion of the GW frequency of the inner orbit) via the differential equation
\begin{equation}
    \dot{u} = \Omega_\out \frac{\left(1+e_\out\cos u\right)^2}{\left(1-e_\out^2\right)^{3/2}},
\end{equation}
where $\Omega_{\out}=\sqrt{M_3/a_\out^3}$. 
In summary, the Doppler shift can be parameterized in terms of 5 parameters: $(\Omega_\out, e_\out, \mathcal{A}, \gamma, u_{\rm c})$ with $u_{\rm c}=u(t=t_{\rm c})$, and our goal here is to illustrate that $\Omega_\out$ and $e_\out$ can both be measured with high accuracy. 

\begin{figure}
  \centering
  \includegraphics[width=0.46\columnwidth]{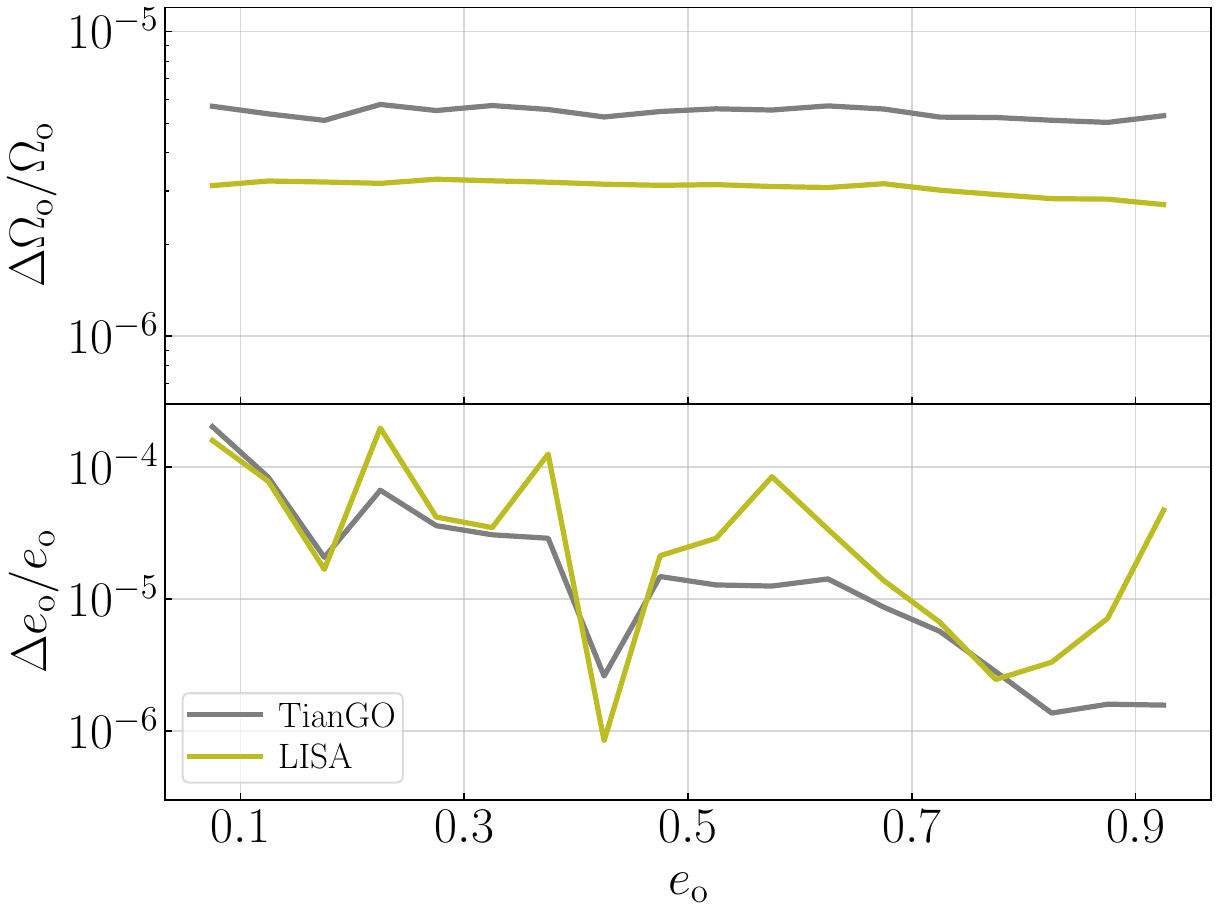}
  \caption{Fractional uncertainties in $\Omega_\out$ (top) and $e_\out$ (bottom) as a function of $e_\out$. The grey (olive) trace assumes the sensitivity of TianGO (LISA). We have dropped other antenna responses and used the angle-averaged sensitivity when evaluating the Fisher matrix ($\sqrt{5}$ times greater than the intrinsic noise). When generating the waveform, we have used $2\pi/\omega_{\out}=0.51\,{\rm yr}$ and $\mathcal{A}=212\,{\rm AU}$, which can be further realized with $M_3=10^8\,M_\odot$, $a_\out=300 M_3$, and $\iota=45^\circ$. }
\label{fig:dOmega_de_ecc_out}
\end{figure}

In Figure~\ref{fig:dOmega_de_ecc_out} we demonstrate the detectability of $\Omega_\out$ (top panel) and $e_\out$ (bottom panel) as a function of $e_\out$ using respectively the sky-averaged sensitivity~\cite{Flanagan:98} of TianGO (grey) and LISA (olive). To model the Doppler phase $\Phi_{\rm D}$, we have further assumed $2\pi/\omega_{\out}=0.51\,{\rm yr}$ and $\mathcal{A}=212\,{\rm AU}$. This set of parameters can be further realized by a physical system with $M_3=10^8\,M_\odot$, $a_\out = 300 M_3$, and $\iota=45^\circ$. For reference, the de Sitter precession period for such a system would be $2\pi/\Omega_{\rm dS}=101\,{\rm yr}$ if $e_\out=0$ and $19\,{\rm yr}$ if $e_\out=0.9$. The values of $\gamma$ and $u_{\rm c}$ are both randomized over when generating the plot. Consistent with the main text, we assumed $M_1=M_2=50\,M_\odot$ and $D_{\rm L}$ for the carrier (in fact, only the chirp mass $\mathcal{M}=44\,M_\odot$ matters as we use the leading-order quadrupole formula for the carrier) and the initial frequency is set to $f^{(0)}=12\,{\rm mHz}$ so that the system merges in $T_{\rm obs}=5\,{\rm yr}$. 

As shown in the plot, the frequency $\Omega_{\out}$ is essentially independent of the eccentricity of the outer orbit $e_\out$ and it can be constrained to a high accuracy of $\Delta \Omega_{\out}/\Omega_\out\sim \text{a few}\times 10^{-6}$ by both TianGO and LISA. The fractional error in $e_\out$ shows more scattering due to the randomness of $\gamma$ and $u_{\rm c}$, yet there is a trend that the fractional error decreases as $e_\out$ increases. Even in the worst cases, we still have $\Delta e_\out/e_\out \lesssim 10^{-4}$. We therefore conclude that both $\Omega_{\out}$ and $e_\out$ can indeed be well constrained from the Doppler phase even for elliptic outer orbits. Once we combine them with the period of the de Sitter precession $\Omega_{\rm dS} \simeq 3M_3 \Omega_{\out}/\left[2a_\out (1-e_\out^2)\right]$ as discussed in the main text, we can therefore simultaneously determine both the mass of the center SMBH $M_3$ and the key properties of the outer orbit $(a_\out, e_\out)$. 

In fact, the de Sitter precession also allows us to infer the inclination angle of the outer orbit and thus $\sin \iota$. As a result, we can also infer $a_\out$ from the amplitude $\mathcal{A}$ which makes the parameter inference even more accurate. Similarly, the precession of the periceter (i.e., $\gamma$) would also provide additional constraints on $(M_3, a_\out, e_\out)$ and enhance the accuracy further. 

\section{SNR of elliptic inner orbits}

We now turn to study the effects of the eccentricity of the inner orbit. 

First, we note that the observed waveform can be modeled as the product $\tilde{h}(f) = \Lambda(f)\tilde{h}_{\rm C}(f)$ with $\Lambda$ the antenna response and $\tilde{h}_{\rm C}$ the antenna-independent waveform of the carrier. We can consequently call parameters that affects only $\Lambda$ the extrinsic parameters (including $M_3$, $a_\out$, $e_\out$, $\lambda$, etc.), and those affecting only $\tilde{h}_{\rm C}$ the intrinsic parameters (including $e_\imag$). 

If we ignore the covariance between different elements, the error of an extrinsic parameter $\Delta \theta^{\rm (ext)}$ scales as
\begin{equation}
    \Delta \theta^{\rm (ext)} \sim \frac{1}{|\partial \tilde{h}/\partial \theta^{\rm (ext)}|} = \frac{1}{|\left[\partial \Lambda/\partial \theta^{\rm  (ext)} \right] \tilde{h}_{\rm c}|}.
\end{equation}
Therefore, we can see an intrinsic parameter such as $e_\imag$ affects the detectability of an extrinsic one (such as $M_3$) mostly through changing the overall signal-to-noise-ratio (SNR). Consequently, we can estimate how the inner eccentricity affects the results we drawn in the main text based on circular orbits by considering its effects on the SNR. 

\begin{figure}
  \centering
  \includegraphics[width=0.46\columnwidth]{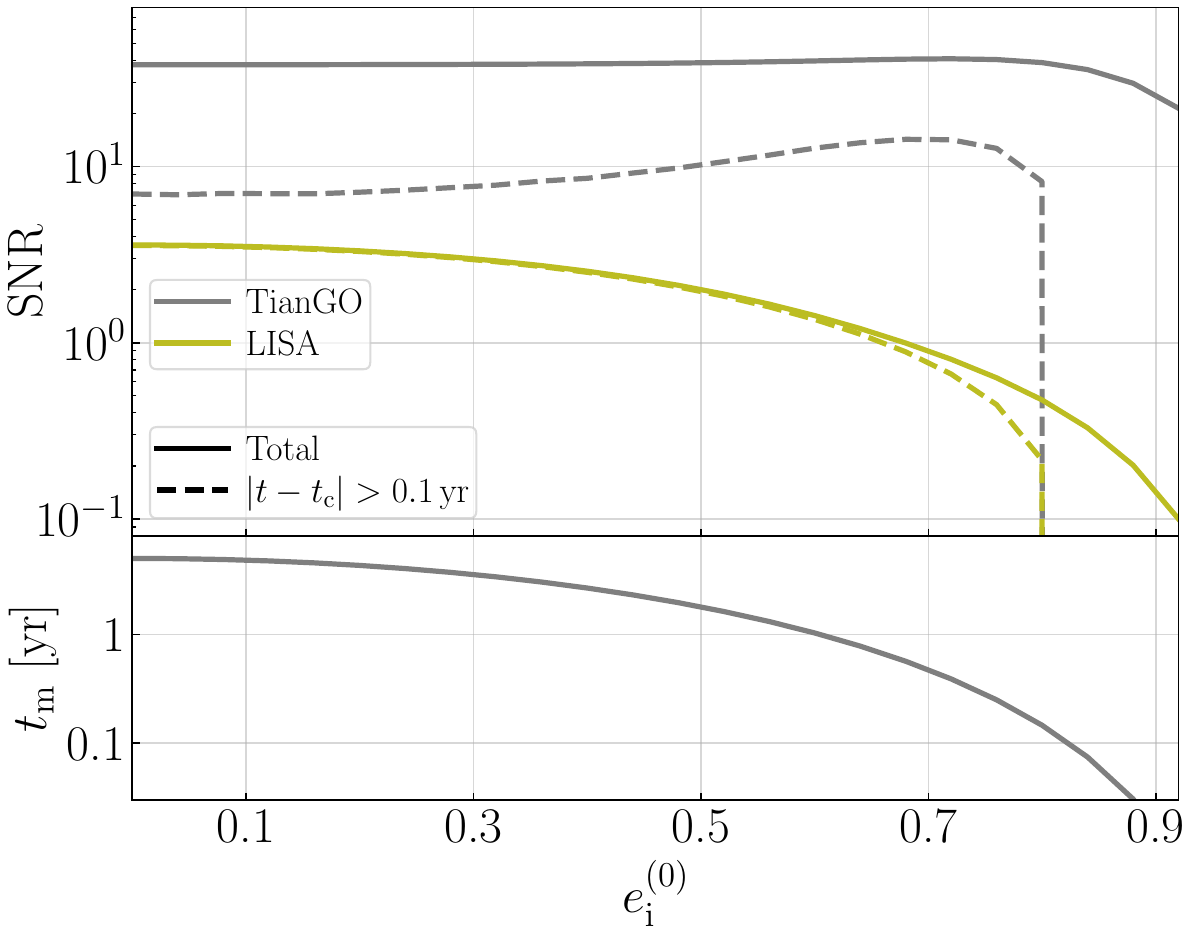}
  \caption{Sky-averaged SNR and merger time as a function of the initial eccentricity of the inner orbit, $e_{\imag}^{(0)}$. The solid trace is the total SNR using all the data and the dashed trace uses only data at least $0.1\,{\rm yr}$ prior to the merger so that this portion is accumulated over a time comparable to the typical precession period. We have fixed the initial semi-major axis of the inner binary to be $a_\imag^{(0)}=1.4\times 10^{-3}\,{\rm AU}$ so that a circular binary can merge in 5 years. If the eccentricity is high ($>0.7$), we would be able to observe the inner binary starting from a much greater $a_\imag$ so that it still stays in band for years with an SNR of few during the early stage evolution (assuming TianGO's sensitivity; see the discussions in the text).}
\label{fig:SNR_vs_ei}
\end{figure}

To estimate the SNR of an elliptic inner orbit, we consider the characteristic strain of the system. Specifically, we can first decompose the time-domain waveform into a sum over harmonics as $h(t)=\sum_k h_k(t)$ with each harmonic oscillating at a frequency $f_k$. Up to corrections due to the precession of the inner pericenter, we have $f_k\simeq k f_\imag$ with $2\pi f_\imag=\sqrt{(M_1+M_2)/a_\imag^3}$. The characteristic strain for each harmonic is thus given by 
\begin{equation}
    h_{{\rm c},k}=\frac{1}{\pi D_{\rm L}}\sqrt{\frac{2\dot{E}_k}{\dot{f}_k}},
\end{equation}
where $\dot{E}_k$ is the GW power radiated to infinity at $f_k$. The SNR can then be obtained by summing over harmonics as 
\begin{equation}
    {\rm SNR}^2=\sum_k\int\frac{h_{{\rm c}, k}^2(f_k)}{5 f_k S_{\rm n}(f_k)} d\ln f_k,
\end{equation}
where $S_{\rm n}$ is the power-spectral density of the instrument noise. Here we drop the antenna response and use the sky-averaged sensitivity instead (which leads to the numerical factor of 5 in the denominator). We refer the interested readers to Ref.~\cite{Barack:04} for more details of the calculation. 

One such example is shown in the top panel of Fig.~\ref{fig:SNR_vs_ei}. Here we consider an inner binary with masses of $M_1=M_2=50\,M_\odot$ and an initial semi-major axis of $a_\imag^{(0)}=1.4\times10^{-3} \,{\rm AU}$ (same as the one considered in the main text). We vary the initial eccentricity $e_\imag^{(0)}$ and compute the SNR based on the characteristic strains using both the sensitivity of TianGO (grey traces) and LISA (olive traces). In addition to the total SNR shown in the solid traces, we also consider the SNR using only the data at least $0.1\,{\rm yr}$ prior to the final merger (shown in the dashed traces). This portion of the data is accumulated over a time comparable to the typical de Sitter precession periods of a few to tens of years and would thus directly helps constraining the time-varying antenna pattern. For reference, we also show the total time to merger $t_{\rm m}$ in the bottom panel.

\begin{figure}
  \centering
  \includegraphics[width=0.46\columnwidth]{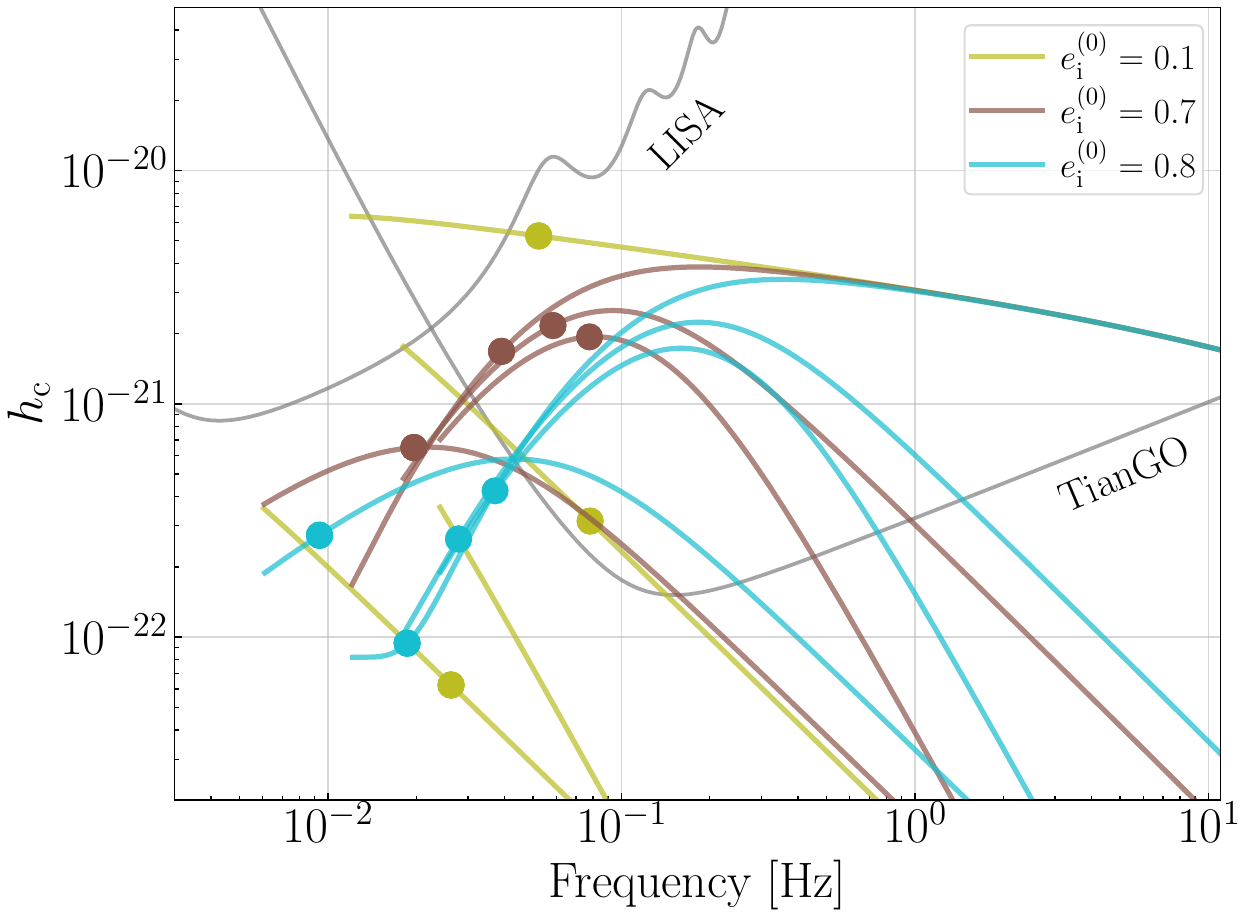}
  \caption{Characteristic strain $h_c$ of the inner orbit for representative initial eccentricities $e_\imag^{(0)}$ (represented by different colors). The initial semi-major axis is fixed to $a_\imag^{(0)}=1.4\times 10^{-3}\,{\rm AU}$. For each configuration we only show the first 4 harmonics. The dot markers correspond to the instant when the inner binary is $0.1\,{\rm yr}$ prior to the final merger.}
\label{fig:hc_vs_ei}
\end{figure}

As shown in the plot, if the inner binary's eccentricity is mild ($e_\imag^{(0)}\lesssim 0.7$; this corresponds to about half of the sources if $e_\imag^{(0)}$ yields a thermal distribution), then both TianGO and LISA see  mild changes in both the total SNR and that accumulated from the early stage only. In fact, the SNR seen by TianGO increases slightly first with an increasing eccentricity. This can be understood by examining Fig.~\ref{fig:hc_vs_ei}. As the eccentricity increases, more GW power are emitted through high-order harmonics (instead of through only the $k=2$ harmonic for circular orbits). These harmonics have higher frequencies and therefore are in the band where a decihertz detector like TianGO is more sensitive. For LISA, the SNR is reduced by a factor of 3 but is still above unity as $e_\imag^{(0)}$ changes from 0 to 0.7. These should not change our results qualitatively. 

When the initial eccentricity is more extreme, $e_\imag^{(0)}\gtrsim 0.8$, the inner binary would merge within $0.1\,{\rm yr}$ and therefore the dashed line vanishes. However, this is an artifact of our fixing $a_\imag^{(0)}=1.4\times10^{-3}\,{\rm AU}$, a value chosen so that our inner binaries (with circular orbits) would merge within 5 years, the fiducial duration of observation $T_{\rm obs}$. In fact, once we allow the inner orbit to be eccentric, we can in fact capture the binary when it is at a much greater orbital separation. For example, an inner binary with $(a_\imag, 1-e_\imag)=(1.4\times10^{-3}\,{\rm AU}, 0.2)$ can be further evolved from $(a_\imag, 1-e_\imag)=(0.05\,{\rm AU}, 6\times10^{-3})$ [note that as shown in Ref.~\cite{Yu:20b}, $(1-e)\propto a^{-1}$ when $(1-e)\ll1$; also note that the Newtonian tide could be important for such an highly eccentric binary with a large semi-major axis, especially if $M_3\lesssim10^7\,M_\odot$] in slightly less than 2 years $<T_{\rm obs}$. During this process, the $k\simeq 3,000$ to $k\simeq 300$ harmonics consecutively sweep through TianGO's most sensitive band, and together they contribute an SNR of 5.3 over the 2-year evolution. As a result, even for significantly eccentric inner binaries, it is still possible to obtain an SNR of a few with an integration time over a year to constrain the time-varying antenna pattern induced by the central SMBH.

\bibliography{ref}